%% file: MainText.tex
\documentclass[
a4paper, 
amsfonts, 
amssymb, 
amsmath,
pra, 
aps, 
showkeys, 
nofootinbib, 
superscriptaddress, 
twoside, 
longbibliography,
twocolumn,
]{revtex4-2}

\input{Preamble/packages}

\makeatletter
\let\old@sverb\@sverb
\def\@sverb#1{\old@sverb{#1}\zz}
\def\zz#1{#1\ifx\@undefined#1\else\penalty\z@\expandafter\zz\fi}
\makeatother

\newcommand{\QuTech}{\affiliation{QuTech, Delft University of Technology, P.O. Box 5046, 2600 GA Delft, The Netherlands}}
\newcommand{\Kavli}{\affiliation{Kavli Institute of Nanoscience, Delft University of Technology, P.O. Box 5046, 2600 GA Delft, The Netherlands}}
\newcommand{\EWI}{\affiliation{Delft Institute of Applied Mathematics, Delft University of Technology, Mekelweg 4, 2628 CD Delft, The Netherlands}}
\newcommand{\TNO}{\affiliation{Netherlands Organisation for Applied Scientific Research (TNO), P.O. Box 96864, 2509 JG The Hague, The Netherlands}}
\newcommand{\SRON}{\altaffiliation[Present address: ]{SRON Netherlands Institute for Space Research, Niels Bohrweg 4, 2333 CA Leiden, The Netherlands}}

\newcommand{\nametitle}{Coherent versus stochastic error injection on a repetition-code logical qubit in superconducting hardware}
\newcommand{\qec}{QEC\xspace}
\newcommand{\dOne}{D_3\xspace}
\newcommand{\dTwo}{D_4\xspace}
\newcommand{\dThree}{D_5\xspace}
\newcommand{\dFour}{D_2\xspace}
\newcommand{\dSeven}{D_1\xspace}
\newcommand{\zThree}{Z_1\xspace}
\newcommand{\zOne}{Z_2\xspace}
\newcommand{\xOne}{Z_3\xspace}
\newcommand{\xTwo}{Z_4\xspace}

\newcommand{\mwpm}{MWPM}

\newcommand{\suppress}{1.45}

\newcommand{\testbed}{testbed\xspace}
\newcommand{\bitflip}{bitflip\xspace}
\newcommand{\codedistance}{$d$\xspace}
\newcommand{\disthree}{$d$=3\xspace}
\newcommand{\disfive}{$d$=5\xspace}
\newcommand{\disthreeandfive}{$d$=3 and $d$=5\xspace}
\newcommand{\disnine}{$d$=9\xspace}

\newcommand{\thetac}{\theta_\mathrm{c}}
\newcommand{\pc}{p_\mathrm{c}}
\newcommand{\zzc}{$ZZ$\xspace}
\newcommand{\wtwoztype}{weight-2 $Z$-type\xspace}
\newcommand{\surfacecode}{surface code\xspace}
\newcommand{\surfacecodes}{surface codes\xspace}
\newcommand{\repetitioncode}{repetition code\xspace}
\newcommand{\repetitioncodes}{repetition codes\xspace}
\newcommand{\densitymatrix}{density-matrix\xspace}
\newcommand{\capdensitymatrix}{Density-matrix\xspace}
\newcommand{\psiA}{\psi_{\rm A}}

\begin{document}

\title{\nametitle}
\author{S.~L.~M.~van~der~Meer}\QuTech\Kavli
\author{M.~Serra-Peralta}\QuTech\EWI
\author{Y.~Xin}\QuTech\Kavli
\author{M.~Finkel}\QuTech\Kavli
\author{H.~M.~Veen}\SRON\QuTech\Kavli
\author{M.~W.~Beekman}\TNO
\author{L.~DiCarlo}\QuTech\Kavli
\author{B.~M.~Terhal}\QuTech\EWI

\date{\today}

\begin{abstract}
The performance of quantum error correction (\qec) codes is limited by the underlying physical noise. Theoretical studies show that coherent and stochastic noise have different effects when performing \qec with either surface or \repetitioncodes. We use the \bitflip \repetitioncode, realized in a transmon quantum processor, as a \testbed to experimentally study the impact of injecting coherent versus stochastic errors on the logical performance. 
We adapt a scalable free-fermion simulator to simulate the experiments and we modify a subset sampling technique to efficiently sample stochastic noise in the quantum circuit. 
In the experiment, we do not observe the difference in logical fidelity predicted by simulation for either the distance-3 or distance-5 repetition codes.
We hypothesize that this discrepancy could be explained by small drifts in qubit frequencies, which introduce phase-coherent noise that `stochastifies' the injected coherent errors. Our work contributes to advancing an understanding of how coherent errors affect experimental \qec.
\end{abstract}
\maketitle

\section{Introduction}

Theoretical and numerical research on stabilizer codes typically focuses on stochastic depolarizing noise, where Pauli errors are applied randomly at certain locations of the circuit, since such circuits are efficiently simulatable via the Gottesman-Knill theorem~\cite{gottesman1997stabilizer, gottesman1998heisenberg}. However, in practice, quantum systems experience much richer noise which cannot be described as the stochastic application of an error, let alone a stochastic Pauli error. To compare the effect of two extremes, namely purely coherent and stochastic Pauli noise, we can consider the following single-qubit channels:
\begin{align} \label{eq:noise_channels}
\begin{split}
    &\mathcal{E}_{\mathrm{stoch}}(\rho) := (1-p)\rho + pX\rho X, \\
    &\mathcal{E}_{\mathrm{coh}}(\rho) := R_X(\theta)\rho R_X(-\theta).
\end{split}
\end{align}
The average fidelities of these channels are 
\begin{equation}
    F_{\mathrm{stoch}} = 1-\frac{2}{3}p \;\;\text{and}\;\; F_{\mathrm  {coh}} = 1-\frac{2}{3}\sin^2(\theta/2),
\end{equation}
leading to the comparative identification $p = \sin^2(\theta/2)$. Any coherent error can be brought to such a stochastic Pauli channel by Pauli twirling~\cite{Bennet1996}. There has been theoretical interest to determine whether this identification gives similar results in practice, or whether one has to use a worst case measure, i.e., the diamond norm~\cite{Kitaev02}, $\eta_{\rm stoch}=||\mathcal{E}_{\mathrm{stoch}}-\mathcal{I}||_{\diamond}$ versus $\eta_{\rm coh}=||\mathcal{E}_{\mathrm{coh}}-\mathcal{I}||_{\diamond}$ to capture the noise strength. Since $\eta_{\rm stoch}\sim p$, while $\eta_{\rm coh}\sim |\theta| \sim \sqrt{p}$ \cite{Sanders15}, coherent errors can have a worse effect than stochastic errors.

Using $p = \sin^2(\theta/2)$, Ref.~\cite{Suzuki17} showed that the threshold for the \repetitioncode is lower for coherent than for stochastic errors, using a minimum-weight perfect-matching (MWPM) decoder~\cite{Dennis02, Wang2003}. These numerical results were obtained by efficiently simulating circuits subjected to circuit-level coherent noise, see a review and discussion in the supplementary material~\cite{Supplmat}.
For the \surfacecode, Ref.~\cite{BEKN:coherent} showed that while the threshold for incoming data-qubit coherent errors remains comparable to the stochastic error model, stochastic models still exhibit better performance below the threshold under MWPM decoding.
To study the fundamental noise tolerance of a code independently of a decoder, one can consider the maximum-likelihood decoding problem to locate the phase transition separating the ordered, correctable phase from the disordered, uncorrectable regime. For a \surfacecode subjected to coherent errors injected at the start of every \qec round, Ref.~\cite{VBB:coherent} argued that this critical transition occurs at an angle $\thetac$, which maps to an effective error probability of $\pc^{\rm coh}\approx 0.18$. Crucially, this theoretical upper bound significantly exceeds the established threshold for purely stochastic depolarizing noise $\pc^{\rm stoch}\approx 0.11$. Although this implies a higher theoretical tolerance for coherent noise, efficient classical decoders like MWPM discard the requisite quantum interference, causing them to systematically fail well below this bound.
Additional free-fermion or quasi-probabilistic Clifford simulation techniques to study coherent or non-Pauli noise for the \surfacecode can be found in Refs.~\cite{Hakkaku21, Bravyi14, Marton23}.

As far as we know, no prior work exists investigating whether noticeable differences exist between coherent and stochastic noise in experimental \qec, although recent theoretical work suggests that differences could be observed in detector firing patterns~\cite{takou2025}. For ease of execution, additional noise injection is commonly done using coherent errors (see e.g. Ref.~\cite{Google25}), neglecting the potential difference with stochastic error injection. 
Here, we use the \bitflip \repetitioncode at distances \disthree and \disfive as a \testbed to investigate the different impact of coherent and stochastic noise injection on the logical error probability in a transmon quantum processor. We will refer to this difference as a coherent-stochastic gap.

Naturally, the impact of injected noise will depend on the device's baseline noise. In a memory experiment, the parity of adjacent data qubits ---corresponding to the stabilizers of the bitflip \repetitioncode--- are repeatedly measured. If there is no baseline noise, the outcome of each parity check remains constant across \qec rounds. When a physical error occurs, the parity outcome changes, which is registered as a defect. The frequency at which these defects occur, known as the detection probability, gives a direct metric of the baseline noise affecting the \qec performance. We calibrate our device by minimizing the detection probability. The baseline noise is quantified using an error budget.
Once this baseline is established, we compare our experimental error injection results against a scalable free-fermion simulation. While our current experiments are sufficiently small to allow for exact \densitymatrix simulations, validating this computationally fast method is important when investigating the coherent-stochastic gap for much larger quantum codes.

Our findings show that, given a (uniform-weight) MWPM decoder, the coherent-stochastic gap is not present in either \disthree or \disfive experiment, while simulation predicts one.
Additional \densitymatrix simulations with more realistic noise models are included in an effort to understand this disagreement. We hypothesize that and numerically test whether the difference between experiment and simulation can be explained by qubit frequency drift induced phase errors.

\section{Coherent and stochastic errors in \qec}
\label{sec:explain_cohstoch}

\begin{figure}
    \centering
    \includegraphics[width=0.9\columnwidth]{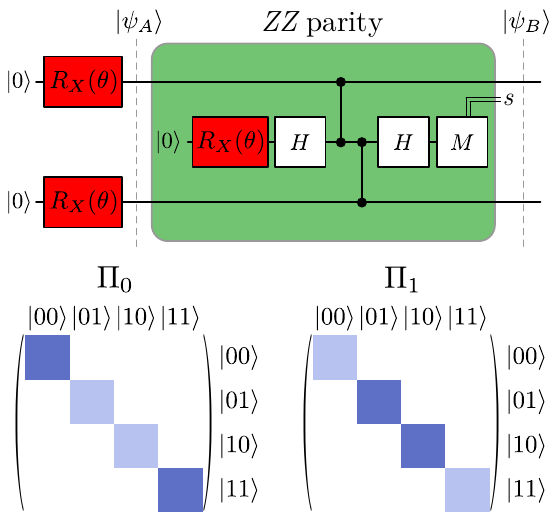}
    \caption{Single \wtwoztype stabilizer used as building block in the \bitflip \repetitioncode memory experiment. The circuit shows injected coherent noise in the red boxes with rotation angle $\theta$ for both data and ancilla qubits. The projectors  $\Pi_s = K_s^{\dagger} K_s$ with $K_s$ in Eq.~\eqref{eq:kraus_s} are depicted below, with the (blue) color intensity being proportional to the matrix element value, estimated when $\theta \ll \pi/2$. }
    \label{fig:coherent_interference}
\end{figure}

A perfect parity-check circuit, which measures the stabilizers of a quantum code, projects any superposition of Pauli errors onto a superposition of errors that share the same syndrome. This projective measurement destroys any quantum coherence between errors associated with different syndromes. 
Within the remaining superposition, errors that share the same syndrome are related either by a product of stabilizers or a logical operation.
Consequently, parity check circuits tend to `stochastify' coherent errors. 
To illustrate this `stochastification', we examine the \wtwoztype parity-check circuit in Fig.~\ref{fig:coherent_interference}. This three-qubit system can be viewed as a minimal error correcting code where the logical operators are $X_L=XX, Z_L=ZI$, and $Y_L=YX$. 

Consider injected $R_X(\theta)$ gates only on the two data qubits. Given the input state $\ket{\psiA}$ which depends on the rotation angle $\theta$, the outgoing state for the no-error outcome $s=0$ is 
\begin{equation}
    \ket{\psi_B} \propto \cos^2\left( \frac{\theta}{2} \right)\ket{00}-\sin^2\left( \frac{\theta}{2} \right)\ket{11} \propto R_{Y_L}(\theta_L) \ket{00},
\end{equation}
which shows that a coherent logical error (instead of a logical \bitflip) can result. However, such a coherent logical error for the $s=0$ case is exponentially suppressed for the distance-$d$ \repetitioncode as $X_L$ has weight $d$. 
Previous work~\cite{Greenbaum_2017, Iverson_2020} suggests that with a stochastic parity-check baseline, injecting coherent versus stochastic errors solely on data qubits results in minimal difference in logical performance. While this observation holds for the standard spacetime MWPM matching graph shown in Fig.~\ref{fig:error_location_performance}(d), such a graph is sub-optimal for data-only error injection. An optimal decoder utilizing only space-like edges reveals a distinct coherent-stochastic gap;
however, this configuration assumes ideal syndrome measurements, whereas experimental syndrome measurements are inherently noisy.
Consequently, we employ a spacetime MWPM decoder through this work to maintain consistency with experimental benchmarks and ensure a uniform comparison across all simulation sets.

In contrast, the dynamics shift when a coherent rotation $R_X(\theta)$ is injected solely onto the ancilla qubit during the \zzc parity check. The resulting Kraus operators for measurement outcomes $s \in \{0, 1\}$, defined as
\begin{equation} \label{eq:kraus_s}
    K_s=\frac{1}{2}(I + (-1)^s e^{i\theta} ZZ),
\end{equation}
with probability $P(s)={\rm Tr}(K_s^{\dagger} K_s \ket{\psiA}\bra{\psiA})$. If we were to replace $R_X(\theta)$ by a stochastic \bitflip, the corresponding map $\mathcal{S}_s$ with outcomes $s=0,1$ still only involves the operator \zzc [Eq.~\eqref{eq:Kraus_parity}]. This implies that any computational basis input state, possibly one with a baseline of stochastic Pauli noise, is left invariant by such parity check measurement. 
Because the coherent rotation $R_X(\theta)$ occurs before the ancilla is read out, it effectively tilts the measurement axes of the parity check. Mathematically, this alters the measurement projectors ($\Pi_0=K_0^{\dagger}K_0$ and $\Pi_1=K_1^{\dagger}K_1$), causing their non-zero matrix elements to spread across different basis states (visualized by the shaded diagonal elements in Fig.~\ref{fig:coherent_interference}). Consequently, the probability of obtaining a specific syndrome outcome $s$ becomes dependent on the injection parameter $\theta$. However, its dependence is identical if the coherent rotation on the ancilla qubit is replaced by equivalent-strength stochastic noise.
For instance, an input state $\ket{00}$ yields the outcome $s=0$ with probability $P(0) = \cos^2(\theta/2)=1-p$, matching the stochastic case. This equivalence extends to the \surfacecode: assuming a baseline of stochastic Pauli noise, coherent ancilla errors acting after state preparation yield dynamics indistinguishable from stochastic ancilla errors\footnote{Note that one could place coherent errors in other locations on the ancilla qubit for which this fact would not be true.}.
We confirm this in Fig.~\ref{fig:error_location_performance}(e), which reveals no observable logical performance gap between coherent and stochastic ancilla-only error injection.

These arguments imply that we can only expect a sizable coherent-stochastic gap if we inject either coherent or stochastic noise on both data and ancilla qubits at the locations in Fig.~\ref{fig:coherent_interference}. In this case, given the input state $\ket{\psiA}$, the unnormalized outgoing state for outcomes $s=0,1$ is
\begin{equation}
\begin{split}
    \ket{\psi_B^s} \propto& \; A_s\left[\cos^2\left( \frac{\theta}{2} \right) \ket{00} - \sin^2\left( \frac{\theta}{2} \right)\ket{11}\right] +\\
    &- A_{s\oplus 1} i\sin\left( \frac{\theta}{2} \right)(\ket{01} + \ket{10}),
\end{split}
\end{equation}
with $A_0 := \cos(\theta/2)$ and $A_1 := -i\sin(\theta/2)$.
Analogous to the case of only data qubit error injection, this can create coherent logical rotations. Crucially, however, imperfect parity mapping also induces state transitions into an orthogonal stabilizer subspace without triggering a syndrome event, thereby potentially worsening the logical qubit performance.
Indeed, Fig.~\ref{fig:error_location_performance}(c) shows that a small gap between injected coherent and stochastic (shaded region) opens up when coherent noise is injected on both data and ancilla qubits, given the stochastic baseline noise. These observations directly inform the error injection strategy for our experimental implementation, where we want to examine the coherent stochastic gap in a typical \qec context with both data and ancilla qubit errors injected.

\section{Experimental implementation}

\begin{figure}[tb]
    \centering
    \includegraphics[width=0.99\columnwidth]{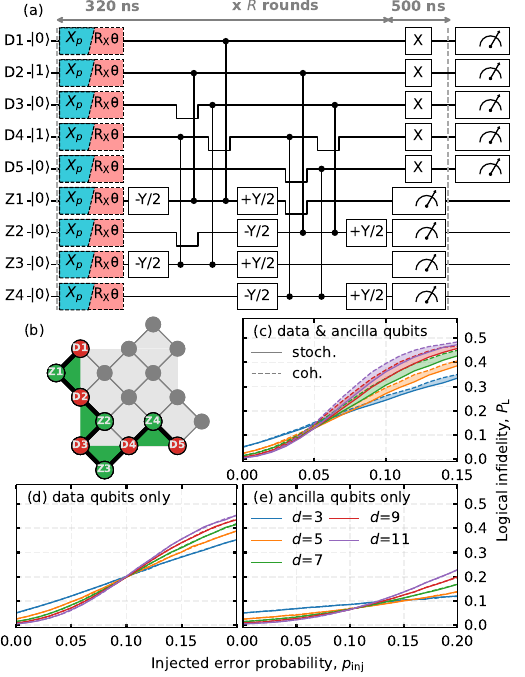}
    \caption{
    Repetition code with error injection circuits and free fermion simulation results. (a) Quantum circuit for the \disfive \bitflip \repetitioncode, showing coherent/stochastic error injection locations at the start of each \qec round. (b)~Code layout on a Surface-17 processor. The qubits $\dFour$-$\dOne$-$\dTwo$ are used for the \disthree code. (c-e) Free-fermion simulations run for $R=d$ \qec rounds, with the stochastic baseline noise specified by $p_{\rm data} = 4\%$ and $p_{\rm anc} = 8\%$, independent of \codedistance. The injected incoming coherent or stochastic noise with corresponding noise threshold is either at data qubit only ($\sim10\%$), ancilla qubit only ($\sim12\%$) or data and ancilla qubit locations ($\sim5\%$). 
    The baseline consists of stochastic phenomenological noise.
    Each point is the average of $10^5$ samples.
    }
    \label{fig:error_location_performance}
\end{figure}

To investigate the difference between coherent and stochastic noise in experiment, one wants to work in the regime where the baseline noise is below threshold, i.e., the exponential suppression factor $\Lambda> 1$~\cite{Google21}. This makes the \repetitioncode a good \testbed as its noise threshold is less stringent as compared to the \surfacecode. \repetitioncodes have been studied extensively and are typically used as an entry-level point to estimate the \surfacecode performance~\cite{Kelly15, putterman2025hardware, Google21}. 

The experiment is executed on a Surface-17 processor designed for \qec~\cite{Versluis17, Marques23, Valles23, Krinner22, Zhao22}. 
While it is possible to implement \repetitioncodes up to \disnine in this 17-qubit processor, we restrict ourselves to \disthree and \disfive. To observe the effect of coherent and stochastic noise, errors of the form $\mathcal{E}_{\mathrm{coh}}$ or $\mathcal{E}_{\mathrm{stoch}}$ in Eq.~\eqref{eq:noise_channels} are injected in a memory experiment [Fig.~\ref{fig:error_location_performance}(a)].

\begin{figure}[tb]
\centering
\includegraphics[width=\columnwidth]{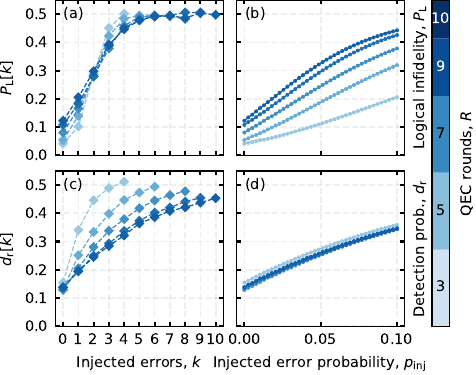}
\caption{Reconstruction of experimental average detection probability and logical error probability (logical infidelity) from the stochastic noise injected circuits of the \disthree \repetitioncode. (a)~Average logical error probability $P_L[k]$ across the circuits with $k$ injected \bitflip errors. (b)~Reconstructed logical error probability $P_L(p_{\rm inj})$ as a function of injected error probability $p_{\rm inj}$. (c)~Average detection probability $d_r[k]$ across the circuits with $k$ injected \bitflip errors. (d)~Reconstructed detection probability as a function of injected error probability $p_{\rm inj}$. Deviations above this value can be explained by possible undersampling of the high number of error configurations $\binom{N_{\rm loc}}{k}$.
} \label{fig:StochasticNoiseStrategy}
\end{figure}
 
From an experimental point of view, the error injection can be done by (1)~intentionally adding gates to the quantum circuit to perform consistent over- or under-rotations~\cite{Google25}, (2)~programming control electronics to interject gate pulses with a random probability in real time~\cite{fruitwala2024hardware} or (3)~pre-compiling the quantum circuit with random gates, a technique mostly recognized from its application in randomized benchmarking~\cite{kim2021hardware}. While coherent errors can be trivially implemented via additional unitary gates, stochastic errors require either real-time pulse injection or pre-compiled randomized circuits. Due to control-hardware limitations that introduce time overhead in real-time error injection [Sec.~\ref{sec:experiment_description}], we use pre-compiled randomized circuits for sampling stochastic noise. This choice offers full control over the sampled injected errors, their location and probability. We cannot use Monte Carlo sampling as a feasible strategy to generate the pre-compiled noisy circuits due to the time overhead it creates from uploading the required amount of unique circuits to our control electronics. 
To solve this issue, we implement a subset sampling technique. By physically executing circuits with a deterministic number $k$ of injected \bitflip errors and classically weighing their results, we can reconstruct the overall logical error probability, $P_L(p_{\rm inj})$, avoiding the overhead of random circuit generation using Monte Carlo sampling. It holds that
\begin{equation} \label{eq:pl_main}
P_L(p_{\rm inj})=\sum_{k=0}^{N_{\rm loc}} A_k(N_{\rm loc},p_{\rm inj}) P_L[k],
\end{equation}
with $P_L[k]$ the average logical error probability across circuits with $k$ deterministically-injected \bitflip errors. Here $A_k$ is the binomial prefactor: 
\begin{equation} 
    A_{k}(N_{\rm loc},\,p_{\rm inj}) = \binom{N_{\rm loc}}{k}p_{\rm inj}^k(1 - p_{\rm inj})^{N_{\rm loc} - k},
    \label{eq:a_prefactor}
\end{equation}
with $N_{\rm loc}$ the number of all possible injection locations in the entire memory experiment circuit. Each circuit is uploaded once and sampled many times to average out the baseline noise. Note that, from a single experimental dataset for $\{P_L[k]\}$, we can obtain $P_L(p_{\rm inj})$ for any $p_{\rm inj}$. The sum in Eq.~\eqref{eq:pl_main} is truncated to further reduce the sampling overhead without sacrificing accuracy. We refer to Sec.~\ref{sec:effsamp} in~\cite{Supplmat} for a detailed description. 
This strategy can also be applied to the detection probabilities, which quantify the rate at which the detectors~\cite{Gidney21} get triggered by the errors [Fig.~\ref{fig:StochasticNoiseStrategy}].

A widely used metric for characterizing experimental logical performance is the logical error rate per \qec round, $\varepsilon_L$~\cite{Google21, Kelly15, Google23, Google25}. In the absence of injected errors (referred to as baseline performance), our device achieves $\varepsilon_L=1.7\%$ and $1.2\%$ for the \disthreeandfive \repetitioncodes when decoded with MWPM, respectively, leading to an exponential suppression factor $\Lambda_{3/5}:=\varepsilon_L^{(3)} / \varepsilon_L^{(5)} \approx\suppress$ [Sec.~\ref{sec:charbackground}]. Nevertheless, we use the logical error probability, $P_L(R,p_{\rm inj})$, given $R$ \qec rounds, to study the coherent-stochastic gap for the following reasons.
First, injecting coherent errors can lead to logical rotations which does not fit a single exponential error rate. Secondly, obtaining $\varepsilon_L$ requires executing the circuit with different number of \qec rounds, which makes the circuit compilation overhead for stochastic noise sampling worse. Therefore, the coherent-stochastic gap is given by 
\begin{align}
   \Delta_L(R,p_{\rm inj})=P_L^{\rm coh}(R, p_{\rm inj}) - P_L^{\rm stoch}(R, p_{\rm inj}),
   \label{eq:gap}
\end{align}
with $p_{\rm inj}$ the injected error probability. The specific choice $R=10$ reflects a tradeoff between maximizing the number of \qec rounds to accumulate sufficient coherent interference and having the logical error probability sufficiently below $1/2$ to observe the coherent-stochastic gap [Sec.~\ref{sec:dec}].

Finally, the memory experiments are decoded using a MWPM decoder with uniform-weights on time- and space-like edges in the decoding graph. The weights are uniform for all values of $p_{\rm inj}$ to be as independent of the type of injected noise as possible, even though this sacrifices logical performance [Sec.~\ref{sec:dec}].

\section{Calibration and Characterization of Baseline Noise}
\label{sec:charbackground}

\begin{figure}[tb]
\centering
\includegraphics[width=\columnwidth]{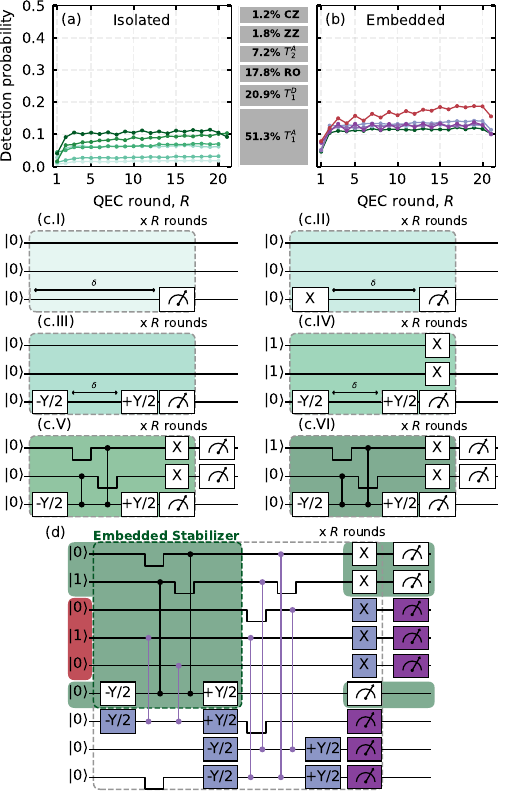}
\caption{
Characterization of effective noise source contributions to an individual stabilizer performance. (a) Experimental stabilizer detection probabilities per characterization circuit (c.I-VI). The experimental data is matched in simulation to produce an error budget (gray bar chart) containing: $T_1^D$/$T_1^A$ (data/ancilla relaxation), $T_2^A$ (ancilla decoherence), RO (readout errors), CZ (two-qubit gate errors) and residual \zzc errors. Each characterization circuit isolates a specific noise contribution from the standard stabilizer circuit (c.VI) used in the \repetitioncode. (b) Experimental stabilizer detection probability in an embedded circuit (d) comparable to the complete \repetitioncode circuit. This circuit evaluates the impact of extended \qec round duration and crosstalk effects on the detection probability. We attribute the jump from $10\%$ to $20\%$ detection probability in (b) to state-dependent errors (red) such as residual \zzc coupling affecting the two-qubit gate performance.
\label{fig:DefectrateCalibration}}
\end{figure}

\begin{figure}[tb]
\centering
\includegraphics[width=\columnwidth]{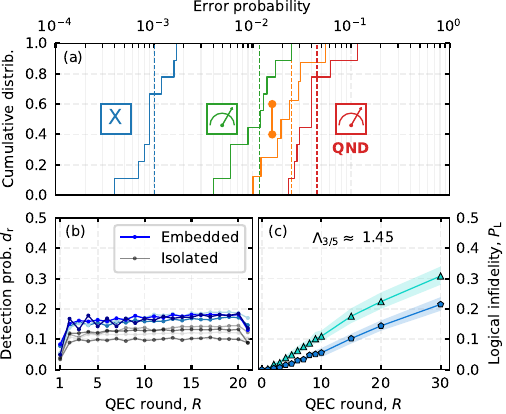}
\caption{
Baseline experimental performance of individual operations, detection probabilities, and logical performance for the \disthreeandfive \bitflip \repetitioncodes. (a) Cumulative distribution of single-qubit gate, CZ gate and readout performance [Tab.~\ref{tab:dm}]. (b) Average detection probability of the four \wtwoztype stabilizers of the \disfive \repetitioncode. One of these stabilizers ($\zOne$) is shown in detail in Fig.~\ref{fig:DefectrateCalibration}(a-b). The gray data represent the individual stabilizer performance in isolation, the colored (blue) data represent the simultaneous (embedded) performance while running the full \repetitioncode. (c) Logical infidelity curves for the memory experiment with \disthree (triangle marker) and \disfive (pentagon marker) \repetitioncodes. The shaded region indicates the minimum and maximum performance bounds over a 2 hour period.
\label{fig:DefectratePerformance}}
\end{figure}

Any experimental quantum processor suffers noise coming from physical processes such as qubit decoherence, thermal excitations, control imperfections, and parameter drifts which constitute a baseline mixture of coherent and stochastic noise sources. The injected errors -- coherent or stochastic -- interfere both with one another and with the inherent baseline noise of the system. We expect that minimizing the baseline noise further below the threshold increases $\Lambda_{3/5}$, enhancing the sensitivity of the logical performance to the injected errors.

After standard gate calibration, we perform a sequence of fine-tuning calibration steps. This routine is specific to \qec: its purpose is to isolate specific noise contributions and measure how they impact the detection probability, see the circuit variants shown in Fig.~\ref{fig:DefectrateCalibration}. 
Each stabilizer is characterized in isolation and in a fully embedded setting.
On the level of isolated stabilizer performance, each characterization circuit in Fig.~\ref{fig:DefectrateCalibration}(c.I-VI) is sensitive to a buildup of noise sources. The contribution to the detection probability compounds sequentially with each circuit variant. Circuits (c.I) through (c.III) isolate ancilla-specific errors: residual excitations and state-dependent readout errors, followed by ancilla relaxation and dephasing. Circuit (c.IV) introduces data-qubit relaxation and residual \zzc coupling between neighboring data and ancilla qubits. Finally, circuits (c.V) and (c.VI) capture the two-qubit gate and parking-related errors for even and odd initial state parities, respectively. The error budget is calculated for the isolated stabilizer case to understand the contribution of each noise category [Sec.~\ref{sec:defcal}]. At the embedded stabilizer level [Fig.~\ref{fig:DefectrateCalibration}(d)], the characterization includes: (AC) flux and microwave crosstalk, multiplexed readout and residual \zzc coupling.

To characterize all-to-all sensitivity during simultaneous flux pulse operations—specifically for CZ gate execution and qubit parking—we measure the system's flux crosstalk. Assuming a linear response, these crosstalk effects are calibrated to suppress two-qubit phase deviations and leakage errors for each gate operation. The compensation is implemented at the pulse-schedule compilation level.
In this experiment, no leakage reset operations~\cite{Marques23, xin25} or leakage post-selection are used.

The average detection probabilities after the described calibrations are 20\% [Fig.~\ref{fig:DefectratePerformance}(b)]. We attribute the increase in detection probability to state-dependent errors such as residual \zzc coupling affecting the two-qubit gate performance when going from isolated to embedded configuration. The exponential suppression factor $\Lambda_{3/5}$ is calculated based on individually optimized \disthreeandfive \repetitioncodes [Fig.~\ref{fig:DefectratePerformance}(c)]~\cite{Chen21}. This ensures a fair comparison: $\varepsilon_L^{(3)}$ is derived from running dedicated \disthree circuits embedded within the processor, rather than classically sub-sampling the \disfive data (which would artificially include physical crosstalk from active neighboring qubits not present in a true \disthree execution). The baseline performance yields $\Lambda_{3/5} > 1$.

\section{Analyzing the effect of coherent and stochastic noise injection}

\begin{figure}[tb]
\centering
\includegraphics[width=\columnwidth]{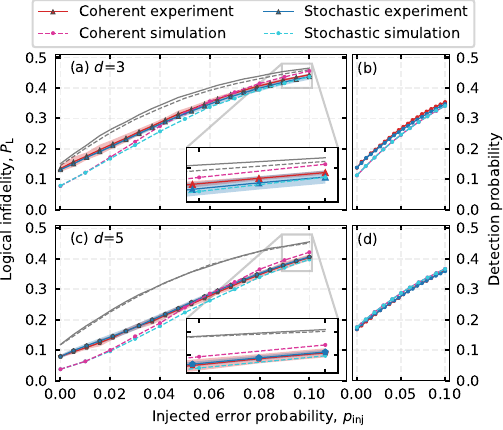}
\caption{
Experimental and free-fermion simulation results of coherent and stochastic noise injection up to $p_{\rm inj}=10\%$ at $R=10$. (a) and (b), Logical infidelity and detection probability for the \disthree \repetitioncode, respectively. (c) and (d), Logical infidelity and detection probability for the \disfive \repetitioncode, respectively. MWPM (red and blue lines) and majority-voting (gray lines) decoders are used to obtain the logical infidelity. The single standard deviation uncertainties are visualized as colored bounds. The number of simulated samples per point is $10^5$. 
\label{fig:ErrorInjectionResults}}
\end{figure}

Two simulators are used to model the experiments: a free-fermion simulator for its scalability, and a \densitymatrix simulator for simulating richer baseline noise~\cite{Obrien17}. We refer to Sec.~\ref{sec:dm_sim} and Sec.~\ref{sec:sps} for a detailed description of the simulators. The efficiency of the free-fermion simulator allows one to reach code distances that are not feasible for \densitymatrix simulations, see Fig.~\ref{fig:error_location_performance}(c-e), at the cost of being able to model limited noise sources. While this efficiency is remarkable---enabling simulations of distance-401 \repetitioncodes on a standard laptop---our immediate priority is to anchor these numerical models against physical reality. Turning to our hardware implementation, the experimental results for coherent and stochastic noise injection in the \repetitioncodes are shown in Fig.~\ref{fig:ErrorInjectionResults}.
While our simulations predict a subtle coherent-stochastic gap of $1-2\%$, resolving this difference experimentally proves challenging. For both \disthree and \disfive, the measurement uncertainties ($\sim 2\%$) are comparable to the expected effect size. Rather than claiming the strict absence of a gap, we hypothesize that unmitigated coherent phase errors effectively 'stochastify' the coherent rotations and obscuring the expected divergence.

\begin{figure}[tb]
\centering
\includegraphics[width=\columnwidth]{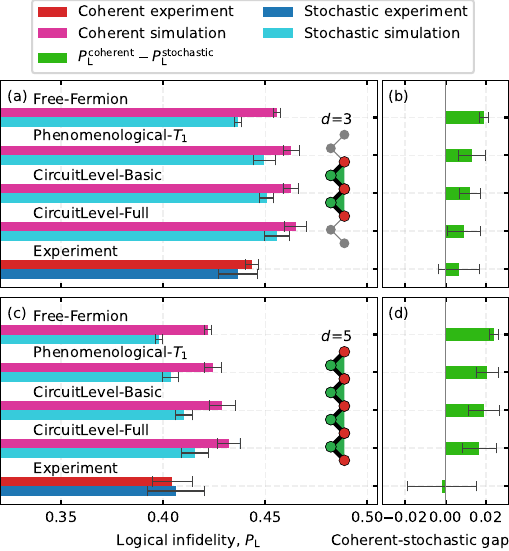}
\caption{
Detailed comparison between free-fermion, \densitymatrix simulations, and experimental results. (a) and (c)~show the discrepancy between simulated and experimental results in logical infidelity at $p_{\rm inj}=10\%$ and $R=10$. The simulation models are ordered from fast and scalable (free-fermion) to realistic and complex (full circuit-level \densitymatrix). (b) and (c) highlight the coherent-stochastic gap. The uncertainties indicate one standard deviation from the mean.
\label{fig:CompareDensityMatrixSimulation}
}
\end{figure}

To unpack the discrepancy between the expected theoretical gap and our experiment, we first benchmark these results against our scalable free-fermion simulations. The baseline noise model used in the simulation is phenomenological, containing (baseline) noise channels, $\mathcal{S}_{\rm BL}$, on both data and ancilla qubits at the beginning of the \qec round. In particular, $\mathcal{S}_{\rm BL}$ corresponds to \bitflip channels to keep the simulation scalable and are determined by noise parameters $p_{\rm data}$ and $p_{\rm anc}$ for data and ancilla qubits, respectively. 

The parameters are adjusted to match the experimental average detection probability and the logical infidelity from majority voting [Sec.~\ref{sec:ff_matching}]. This can be done independently of our choice of MWPM decoding. Given this baseline noise model, the free-fermion simulation predicts a small coherent-stochastic gap for both the \disthreeandfive \repetitioncodes.

The difference between free-fermion simulation and experiment at low injected error probability is most likely caused by the baseline noise model approximations performed in the free-fermion simulator, that is: (1)~using a phenomenological \bitflip noise instead of a general circuit-level noise, where each operation is considered noisy, and (2)~not using qubit-specific noise parameters. The simulation has better agreement with the experiment at high injected error probability, where the injected noise dominates. Note that even if we match the experimental baseline performance ($p_{\rm inj} = 0\%$) by increasing $p_{\rm data}$ and $p_{\rm anc}$, the free-fermion simulator still predicts a coherent-stochastic gap [Sec.~\ref{sec:ff_matching}], showing that the difference in gap size is not caused by underestimating the experimental baseline noise.

In an effort to understand the disagreement between experiment and the free-fermion simulation, we employ a \densitymatrix simulator to test a more realistic noise model by including e.g., decoherence and leakage [Fig.~\ref{fig:CompareDensityMatrixSimulation}]. The first model (Phenomenological-$T_1$) maintains a phenomenological baseline but replaces simple \bitflip channels with physical amplitude damping. To ensure a fair comparison to the free-fermion simulation, the \bitflip error probability $p$ and the amplitude damping parameters are related by
\begin{equation}
    p=\frac{1}{2}\left(1-e^{-t/T_1}\right),
    \label{eq:defp}
\end{equation}
with $T_1$ the relaxation time and $t$ the time during which the qubit can decay. 

The subsequent models introduce full circuit-level noise: the CircuitLevel-Basic version incorporates relaxation, dephasing and readout errors, while the CircuitLevel-Full version further extends the baseline noise model by simulating transmons as qutrits, including coherent evolution of the $\ket{f}$ state, leaky CZ gates, two-qubit phase errors, and residual \zzc coupling. Regarding the choice of noise parameters, we use the values obtained from experimental characterization except for the relaxation and dephasing times, $T_1$ and $T_2$, which are manually adjust to match the experimental average detection probability and the majority voting performance as reference. We allow tuning these two parameters as each simulation has a different amount of noise, thus requiring compensation. Section~\ref{sec:dm_sim} describes the modeling of the operations, noise and how parameters are chosen to match the experiments. 

\begin{figure}[tb]
\centering
\includegraphics[width=\columnwidth]{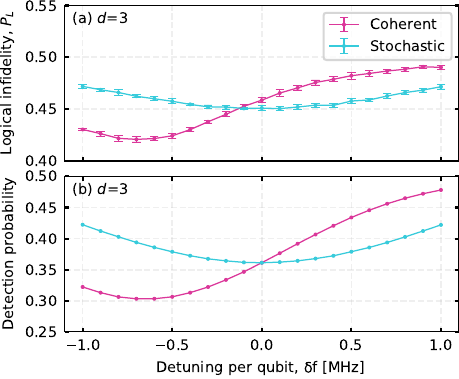}
\caption{
Impact of coherent phase errors on \repetitioncode performance. (a) \capdensitymatrix simulation (CircuitLevel-Basic) of the logical infidelity with coherent (red) and stochastic (blue) $X$ error injection on a \disthree \repetitioncode (at $p_{\rm inj}=10\%$ and $R=10$) as a function of injected phase error $R_Z(\phi)$. (b) The corresponding detection probability. The phase error is applied as circuit-level noise on both data and acilla qubits.
Coherent phase errors that are introduced while the ancilla qubits are in the XY-plane (during CZ gates) are translated to $R_X(\phi)$ rotations and interfere with the injected coherent errors. Therefore, the detection probability and logical infidelity is asymmetrically affected depending on the sing of the incoming phase error.
The simulated number of samples per point is $5\cdot 10^5$.}

\label{fig:CoherentOscillation}
\end{figure}

For \disthree, the \densitymatrix simulations show that the coherent-stochastic gap gets reduced when including more realistic error sources to the point where it is comparable to experiment, see Fig.~\ref{fig:CompareDensityMatrixSimulation}(b). However, while the \disfive code shows no coherent-stochastic gap in experiment, both the free-fermion and \densitymatrix simulations predict a $\sim 2\%$ gap. Similarly to the \disthree case, we observe a slight decrease in the gap when increasing the noise complexity in the \densitymatrix simulations.
This suggests that the true baseline noise acts as a natural dephasing or twirling mechanism, effectively turning coherent errors into stochastic errors before they can interfere.

We believe that the absence of a coherent-stochastic gap in the experiment may be due to extra noise mechanisms that we have not considered in simulation. 
We investigate whether the discrepancy can be explained by the accumulation of (coherent) phase errors during the \qec rounds. In particular, we focus on
the effect of drift in the qubit frequencies, which leads to a detuning between the qubit and drive reference frames, resulting in an accumulating phase error over time~\cite{Klimov18, schlor2019correlating}.
Other potential sources include: qubit frequency calibration errors, single-qubit phase rotations produced by two-qubit gates and parking which we corrected for using virtual phase updates, and readout induced dephasing which we address with dynamical decoupling ($X$) gates.

While the logical observable of the bitflip \repetitioncode is not directly sensitive phaseflip errors, the presence of coherent phase ($R_Z(\phi)$) and rotation ($R_{X}(\theta)$) errors do affect the expectation value of the logical observable~\cite{kurilovich2025correlated}.
Specifically, a shift in the qubit transition frequency by $\delta f$ introduces a longitudinal term to the Hamiltonian in the rotating frame:
\begin{equation}
    H_{rot} = -\frac{\hbar}{2}(2 \pi \delta f) Z.
\end{equation}
This term causes the qubit state to undergo a deterministic precession around the $z$-axis of the rotating frame. For simplicity, we assume that $|\delta f|$ is much smaller than the bandwidth of the drive pulses, rendering the off-resonance error during active gates negligible. However, during idle periods of duration $t$, this detuning accumulates as a phase error $\phi = 2 \pi \delta f t$, which can be modeled via circuit-level $R_Z(\phi)$ operations every QEC round. In the quasi-static limit, where the noise coherence time is much longer than a single experimental run, $\delta f$ is treated as constant for the duration of the circuit.

To test this hypothesis, we repeated the CircuitLevel-Basic simulations but now including circuit-level phase rotations on data and ancilla qubits as a function of drive detuning per qubit, see Fig.~\ref{fig:CoherentOscillation}. 
Our results indicate that both logical infidelity and detection probability are sensitive to accumulated phase errors under both coherent and stochastic noise regimes. In the stochastic case, however, these metrics increase symmetrically with respect to the qubit detuning sign. This behavior is consistent with the expectation that stochastic and phase noise constitute independent error channels whose contributions additively degrade system performance.

Interestingly, the combination of coherent rotations and phase errors reduce the logical infidelity and detection probability under certain conditions.
The reasoning behind this is that during the CZ-gates, when the ancilla qubit is in XY-plane, the accumulated phase error can interfere destructively with the injected coherent rotation, thereby suppressing the total error.
The minimum in Fig.~\ref{fig:CoherentOscillation} is thus realized when the coherent rotation angle $\theta$ and the accumulated phase error $\phi$ are equal in magnitude and opposite in sign ($\theta = -\phi$).
The distinct ways in which coherent and stochastic noise interact with phase errors can account for the closure of the coherent-stochastic gap and further suggest the possibility of a negative gap. However, in practical implementations, qubits do not experience the same frequency detuning $\delta f$ throughout the duration of the experiment.
Instead, temporal fluctuations in frequency will result in an ensemble average across the simulated response curve shown in Fig.~\ref{fig:CoherentOscillation}.

\section{Conclusion}
In this work, we established a comprehensive experimental and numerical \testbed to evaluate the impact of coherent versus stochastic noise on \repetitioncodes. To overcome control-hardware compilation limits, we adapted a subset sampling technique, enabling an efficient, direct experimental comparison between coherent and stochastic noise types.
Although our simulations predict a subtle coherent-stochastic gap of $1-2\%$ for the \disthree and \disfive versions of the experiment, resolving this difference experimentally proves challenging.
Neither the scalable free-fermion model nor the realistic \densitymatrix simulations fully account for this discrepancy. The \densitymatrix simulations generally predict a smaller gap compared to the idealized free-fermion model, which we attribute to the baseline noise effectively twirling the coherent errors.
We hypothesize that the remaining deviation arises from residual coherent phase errors, such as those induced by qubit frequency drifts over time, which interact differently with the injected coherent and stochastic errors. While this mechanism has been numerically verified, it remains to be experimentally tested.
Because bit-flip \repetitioncodes are distance-1 sensitive to phase errors, they are inherently vulnerable to this unmitigated dephasing. Consequently, addressing these phase errors, either by tracking frequency drifts between experimental runs or by improving baseline qubit coherence, will be crucial for future realizations of this experiment. Finally, extending this framework to \surfacecodes, which provide concurrent topological protection against phase errors, may offer a more robust platform for resolving the coherent-stochastic gap.

\section{Data availability}
The data for all figures in the main text and the supplement are available online at \verb|https://doi.org/10.4121/886d8394-adfe-4560-807c-c0fa007110df|.
The free-fermion simulator is available online at \verb|http://github.com/MarcSerraPeralta/fgs_simulator|.
The \densitymatrix simulator is available online at \verb|https://github.com/MarcSerraPeralta/quantumsim_errorinjection|.

\section{Acknowledgments}
We thank M. H. Shaw for discussions, N. Haider for help in chip design, and S. Vall{\'e}s-Sanclemente, Qblox and Orange Quantum System for technical contributions.
We disclose the use of large language models for support in small mathematical questions, data visualization and literature search.
This work is supported by QuTech NWO funding 2020-2028 – Part~I “Fundamental Research”, project number 601.QT.001-1, financed by the Dutch Research Council (NWO), and the OpenSuperQPlus100 project (no. 101113946) of the EU Flagship on Quantum Technology (HORIZON-CL4-2022-QUANTUM-01-SGA).

\section{Author contributions}
SLMM conceptualized the experiments. SLMM performed the experiment with contributions from YX. SLMM performed the data analysis. SLMM and MSP set up the \densitymatrix simulations. MSP performed scalable simulations and aided with decoding. SLMM, MSP and BMT wrote the manuscript with input from all authors. MB and LDC designed the device. MF and HMV fabricated the device. BMT proposed and supervised the project. LDC supervised the experiment.

\section{Conflicts of interests}
The authors declare no conflict of interests.

\input{Bibliography/bibliography_main}

\clearpage

\renewcommand{\theequation}{S\arabic{equation}}
\renewcommand{\thefigure}{S\arabic{figure}}
\renewcommand{\thetable}{S\arabic{table}}
\renewcommand{\bibnumfmt}[1]{[S#1]}
\renewcommand{\citenumfont}[1]{S#1}
\renewcommand{\thesection}{S\Roman{section}}
\setcounter{figure}{0}
\setcounter{equation}{0}
\setcounter{table}{0}

\onecolumngrid
\section*{Supplementary material for ``\nametitle''}
\FloatBarrier
\twocolumngrid

\setcounter{table}{0}
\renewcommand{\thetable}{S\arabic{table}}
\renewcommand{\theHtable}{S\arabic{table}}
\setcounter{figure}{0}
\renewcommand{\thefigure}{S\arabic{figure}}
\renewcommand{\theHfigure}{S\arabic{figure}}
\setcounter{section}{0}

This supplementary material provides additional information supporting the claims and figures of the main text.

\section{Experimental setup}

The device used consists of 17 flux-tunable transmon qubits in a \disthree rotated \surfacecode layout. All nearest-neighbor transmons are connected through fixed-frequency bus resonators. Every transmon has a dedicated microwave drive and flux-control line, as well as a dedicated pair of readout Purcell-filter resonators. The device hosts three feedlines used for simultaneous multiplexed readout. For a full description of the device layout, see Ref.~\cite{Valles23}. All control and readout signals are generated and acquired at room temperature using a Qblox cluster. Microwave control, baseband flux control, and readout are facilitated by 9 QCM-RF-II modules, 5 QCM modules, and 4 QRM-RF modules, respectively.

\begin{table*}[t]
\label{tab:dm}
\begin{ruledtabular}
\begin{tabular}{lccccccccc}
Single-qubit metrics & $\dSeven$ & $\dFour$ & $\dOne$ & $\dTwo$ & $\dThree$ & $\zThree$ & $\zOne$ & $\xOne$ & $\xTwo$ 				\\
\colrule
Transmon frequency, $\omega_{ge}/2\pi$ (GHz) & 4.788 & 6.802 & 4.843 & 4.788 & 4.907 & 6.032 & 6.087 & 5.891 & 6.107 \\
Anharmonicity, $\alpha/2\pi$ (MHz) & -311 & -299 & -314 & -317 & -309 & -291 & -302 & -302 & -302 \\
Relaxation time, $T_1$ ($\mu$s) & 29.5 & 20.3 & 28.5 & 17.6 & 24.3 & 14.2 & 14.5 & 35.2 & 16.5 \\ 
Ramsey dephasing time, $T_2^*$ ($\mu$s) & 13.5 & 20.2 & 3.3 & 8.3 & 22.7 & 6.5 & 15.5 & 20.1 & 10.2 \\ 
Pure dephasing time, $T_\phi$ ($\mu$s) & 17.5 & 40.2 & 3.5 & 10.9 & 42.6 & 8.4 & 33.3 & 28.1 & 14.8 \\
Average qubit assignment fidelity  (\%) & 99.1 & 98.6 & 98.9 & 97.6 & 98.2 & 98.8 & 98.8 & 99.4 & 98.0 \\
Average QND fidelity (\%) & 97.2 & 95.9 & 95.7 & 88.8 & 95.7 & 96.5 & 96.4 & 98.0 & 90.1 \\
\colrule
Two-qubit metrics & $\zThree\dSeven$ & $\zThree\dFour$ & $\zOne\dFour$ & $\zOne\dOne$ & $\xOne\dOne$ & $\xOne\dTwo$ & $\xTwo\dTwo$ & $\xTwo\dThree$ & $\zOne\dTwo$ \\
\colrule
Residual \zzc coupling, $\zeta_{ZZ}$ (kHz) & -84.4 & -272.6 & -304.2 & -86.9  & -125.3 & -108.0 & -50.0 & -96.3 & -97.0\\
CZ IRB, $\epsilon_{\ket{00}}^{\rm interleaved}$ (\%) & 1.9 & 0.4 & 1.1 & 1.2 & 1.5 & 1.8 & 0.3 & 0.7 & -- \\
Leakage rate, $L_1$ (\%) & $1.2\pm0.1$ & $0.6\pm0.1$ & $0.7\pm0.1$ & $0.6\pm0.1$ & $0.5\pm0.1$ & $1.1\pm0.2$ & $0.8\pm0.1$ & $1.0\pm0.2$ & --\\
\colrule
Operation in \qec experiments & \multicolumn{9}{c}{Duration (ns)} \\
\colrule
Single-qubit gate & \multicolumn{9}{c}{20} \\
Two-qubit gate    & \multicolumn{9}{c}{60} \\
Measurement       & \multicolumn{9}{c}{500} \\
\end{tabular}
\end{ruledtabular}
\caption{Experimentally characterized device metrics, including single-qubit, two-qubit and operation duration metrics. The single- and two-qubit metrics are individually characterized. All qubits are operated at their first-order flux insensitive bias point (sweetspot). Note the two-qubit residual \zzc coupling metric is measured at the idle condition, not including frequency excursions due to CZ-gate or parking operations. Note that an extra residual \zzc coupling term is included for qubit pair $\zOne\dTwo$, these are nearest neighbor qubits and their contribution is included in simulations. Note that the two-qubit gate leakage rate is calculated by assuming all interleaved randomized benchmarking (IRB) leakage originates from the CZ gate. The single-qubit gates are simulated to be ideal.}
\end{table*}

\section{Error injection experiment}
\label{sec:experiment_description}

The \repetitioncode is used to measure the impact of coherent and stochastic noise on the logical fidelity. Since the experiment is designed to mimic phenomenological noise, both coherent and stochastic noise are injected through an additional layer of gates at the start of each \qec round. Single-qubit coherent errors are implemented by adding additional unitary gates ($R_X(\theta)$). Stochastic errors, however, are not native operations and therefore require additional steps to be included in the experiment. 

A typical quantum experiment is designed around a single circuit, executed many times to obtain statistics. This allows control hardware to prioritize acquisition memory over waveform memory and instruction-set memory. The hardware requirements for an experiment that uses several circuits with randomly applied gates, like Monte Carlo sampling of stochastic injected noise, is poorly suited to this design.
To address this limitation while ensuring sufficient sampling, we have employed the technique of subset sampling~\cite{gutierrez2019transversality} and adapted it to include the baseline noise [Sec.~\ref{sec:effsamp}]. 
Alternatively, one could utilize the onboard FPGA of the microwave control modules and program a real-time random number generator (RNG) sampler to randomly execute $X$ gates to inject stochastic errors during the \qec experiment. This approach is ideal regarding circuit compilation, as only one circuit needs to be compiled and uploaded to the control electronics. However, due to technical limitations in our control electronics, there is an 80~ns time overhead in processing and executing the randomly sampled single-qubit gates. This would quadruple our 20~ns single-qubit gate times, extending the total \qec round time by 180~ns. We decided to forfeit this compromised version of real-time error injection approach in favor of a shorter \qec round time.

\subsection{Experiment order and wall clock time}
\label{sec:exp_order_time}

The experiment described in Fig.~\ref{fig:ErrorInjectionResults} consists of a 30-hour measurement window initiated by a full device calibration and organized into ten interleaved iterations of coherent and stochastic error injection batches to ensure that both are evaluated against the same baseline noise fluctuations. The initial calibration consists of single- and two-qubit phase calibrations and frequency optimization to avoid spurious two-level systems (TLS). Each coherent batch requires approximately \qty{1}{\hour} and involves $R=10$ QEC rounds with the injection probability $p_{\rm inj}$ swept from \qty{0.0} to \qty{0.1} in 20 discrete steps, where each configuration is repeated \qty{70000} times and supplemented by readout calibration points for each value of $p_{\rm inj}$ to maintain state assignment accuracy. The stochastic batches, requiring approximately \qty{2}{hours} each due to the computational overhead of unique circuit compilation and waveform uploading to the Qblox cluster, samples \qty{1000} unique circuit realizations for each error weight $k \in \{1, \dots, 10\}$, with \qty{200} shots per circuit to average baseline experimental noise. To monitor whether any TLS moves in frequency and affects any two-qubit gate performance, we perform conditional oscillation experiments for all two-qubit gates used in the experiment which keeps track of single- and two-qubit phases as well as missing fractions, indicating leakage or population exchange with nearby TLS. These conditional oscillation experiments are performed between every batch.

\subsection{Efficient sampling of injected stochastic noise}
\label{sec:effsamp}

When performing Monte Carlo sampling, common stochastic error configurations are sampled several times. This implies that the circuit corresponding to a common configuration of injected stochastic errors is uploaded several times to the control electronics. This is inefficient. Equivalently, the configuration could be uploaded once and included in logical infidelity estimates using a weighing factor [Eqs.~\eqref{eq:a_prefactor-SI} and~\eqref{eq:pl-SI}] representing its probability of occurring; this achieves faster sampling by avoiding unnecessary circuit uploads. This is the idea behind subset sampling~\cite{Heussen24}, which minimizes the number of circuit uploads and allows reusing experimental data for any injected probability by updating the weights.
However, the method faces a fundamental scaling bottleneck: the number of unique error configurations grows exponentially with the number of qubits and the circuit depth. For phenomenological noise injection, the number of error locations in our circuits is $N_{\rm loc} = R \times (2d-1)$, where $R$ is the number of rounds. Even for \disthree and $R=10$, there are more than $10^{15}$ configurations, making it intractable to execute all possible circuits. Consequently, we apply additional approximations to reduce the number of required circuit executions.

The probability for injecting $k$ \bitflip errors across $N_{\rm loc}$ locations follows the binomial distribution,
\begin{equation} 
    A_{k}(N_{\rm loc},\,p_{\rm inj}) = \binom{N_{\rm loc}}{k}p_{\rm inj}^k(1 - p_{\rm inj})^{N_{\rm loc} - k},
    \label{eq:a_prefactor-SI}
\end{equation}
where $p_{\rm inj}$ is the individual error injection probability. Hence, the logical error probability $P_L$ at a given $p_{\rm inj}$ is written as
\begin{align}
P_L(p_{\rm inj})=\sum_{k=0}^{N_{\rm loc}} A_k(N_{\rm loc},p_{\rm inj}) P_L[k],
\label{eq:pl-SI}
\end{align}
where $P_L[k]$ is the average logical error probability across the circuits with $k$ injected \bitflip errors. Truncating the sum to a certain $k_{\rm cutoff}$ will lead to a difference w.r.t. the exact $P_L(p_{\rm inj})$ of at most
\begin{equation} \label{eq:kcutoff}
    \sum_{k=k_{\rm cutoff} + 1}^{N_{\rm loc}} A_k(N_{\rm loc},p_{\rm inj}),
\end{equation}

because $0 \leq P_L[k] \leq 1 \;\forall k$.
We choose $k_{\rm cutoff} = \mu + 2 \sigma$, with $\mu = p_{\rm inj}N_{\rm loc}$ and $\sigma = \sqrt{p_{\rm inj}(1-p_{\rm inj})N_{\rm loc}}$.
When $P_L[k_{\rm cutoff}]$ approaches $1/2$, we further refine the estimate of the logical error probability using the approximation
\begin{equation}
\begin{split}
    P_L(p_{\rm inj})\approx &\sum_{k=0}^{k_{\rm cutoff}} A_k(N_{\rm loc},p_{\rm inj}) P_L[k] \\ &+\sum_{k = k_{\rm cutoff} + 1}^{N_{\rm loc}} A_k(N_{\rm loc},p_{\rm inj}) \frac{1}{2}.
\end{split}
\label{eq:pl_approx-SI}
\end{equation}
The detection probabilities $d_r$ are obtained following the described procedure but changing $P_L$ to $d_r$.

For \disthree and $R=10$ at the highest injection probability $p=10\%$, the cut-off point is $k_{\rm cutoff} = 10$, ensuring $k_{\rm cutoff} \ll N_{\rm loc} = 50$.
In order to estimate $P_L[k]$, we randomly generate $N_k$ circuits with $k$ \bitflip errors out of the large, ${N_{\rm loc}\choose k}$, possible choices following a uniform distribution. 
Each circuit is executed $N_{\rm shot}$ times to ensure a robust average of the baseline noise. For $k \in [2, k_{\rm cutoff}]$, we utilize $N_{\rm shot} = 200$ and $N_k = 1\,000$. For $k=0$ and 1, as ${N_{\rm loc}\choose k} < N_k$, we have generated all possible combinations and enforce the same total sample size $N_k \cdot N_{\rm shot}$ by setting $N_{\rm shot}^{k=1, 2} = \lceil N_k \cdot N_{\rm shot} / {N_{\rm loc}\choose k} \rceil$. Although $N_k$ can be very small compared to ${N_{\rm loc}\choose k}$, free-fermion simulations show that $P_L[k]$ does not vary significantly when using different sets of $N_k$ circuits. In particular, for \disfive, the maximum spread for $P_L[k]$ is $\sim 0.5\%$.

The experimental results in Fig.~\ref{fig:StochasticNoiseStrategy}(a) show that the logical error probability reaches $P_L[k] = 1/2$ when injecting $k=7$ errors ($R=10$). We expect $P_L[k]$ to be well approximated by 1/2 when injecting $7 \leq k \leq k_{\rm cutoff}$, justifying the use of Eq.~\eqref{eq:pl_approx-SI}.
While setting $k_{\rm cutoff} = 7$ would suffice for $P_L(p_{\rm inj})$, this does not hold for the detection probabilities [Fig.~\ref{fig:StochasticNoiseStrategy}(c)], where $d_r[10] \approx 0.45$. Consequently, we maintain $k_{\rm cutoff} = 10$ to ensure consistency across all metrics. Due to the large number of contributions to the uncertainty of $P_L$, we employ bootstrapping to compute the confidence intervals.

Note that to minimize the uncertainty of $P_L$, the best choice of $N_{\rm shot}$ and $N_k$ is such that the uncertainty of $P_L[k]$ is proportional to $1 /(A_k(N_{\rm loc}, p_{\rm inj}))$~\cite{Heussen24}. Our choice of $N_{\rm shot}$ and $N_k$ is not optimal in this sense but allows for a faster sampling given the technical limitations of the control electronics.

\subsection{Decoder}
\label{sec:dec}

\begin{figure}
    \centering
    \includegraphics[width=0.99\columnwidth]{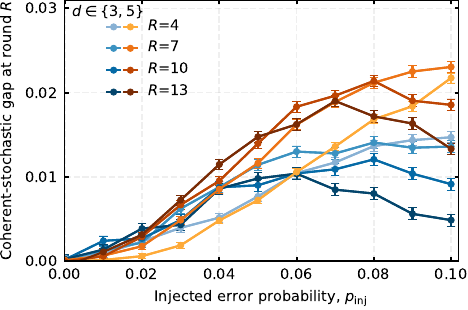}
    \caption{Free-fermion simulation of the coherent-stochastic gap for a \disfive \bitflip \repetitioncode with injected phenomenological stochastic or coherent noise. The memory experiments are run for $R$ rounds and $10^6$ samples, with stochastic phenomenological baseline noise matching the experimental performance ($p_{\rm data} = 3.3\%$ and $p_{\rm anc} = 7.3\%$), see Fig.~\ref{fig:fgs_simulator}(a) for the circuit and noise channels for \disthree. The selected noise parameters are close to the ones which are used for the experimental fit [Sec.~\ref{sec:ff_matching}]. The detection probabilities go from $\sim 20\%$ at no injected errors to $\sim 40\%$ at $p_{\rm inj}=0.1$. }
    \label{fig:gap_vs_rounds}
\end{figure}

This work uses a minimum weight perfect matching (\mwpm) decoder with a uniformly-weighted decoding graph consisting of only space- and time-like edges, and no space-time like edges (see e.g. Ref.~\cite{Kelly15, ali_2024} for \repetitioncode decoding with all three types of edges or more). These edges included in the decoding graph correspond to errors which occur in phenomenological noise, that is: data-qubit $X$ errors before each \qec round leading to space-like edges, and ancilla-qubit measurement errors leading to time-like edges. We have not included space-time edges coming from circuit-level noise because their weights would be unrealistically high (when chosen to be the same as the weight of a space- or time-like edge). The choice of uniform weights is based on not knowing the best performing weights for decoding coherent errors, as MWPM is not even the best decoding strategy for this type of errors~\cite{VBB:coherent, pato2024optimal}. When injecting stochastic errors, the optimal choice would be to update the edge weights based on the injected error probability $p_{\rm inj}$. However, this may not be the optimal choice for decoding coherent errors, thus creating a bias towards better logical performance when injecting stochastic noise. 
Therefore, decoding is performed with uniform weights at all injection probabilities $p_{\rm inj}$. This choice also simplifies data analysis as the same decoder and decoding graph must be used to obtain $P_L[k]$ for a choice of $k$ to ensure that we are fairly estimating $P_L$. 

The decoder has access to the detector information at each round (following the Stim nomenclature~\cite{Gidney21}), with the final data-qubit measurement outcomes incorporated in the detectors representing an effective $R+1$ round. We use \verb|PyMatching|~\cite{higgott2021pymatching} for the implementation of the MWPM decoder. The majority-voting decoder returns the minimum Pauli-weight $X$-type error on the data-qubits that returns the final data-qubit measurement outcomes to the initial codespace, determined by the initial bitstring. The logical infidelity is assessed at round $R=10$ because its coherent-stochastic gap is estimated as one of the biggest compared to other rounds across all evaluated $p_{\rm inj} \in [0, 0.1]$, see Fig.~\ref{fig:gap_vs_rounds}. For $R\rightarrow \infty$, the logical error probability or infidelity $P_L(p_{\rm inj})\rightarrow 1/2$ for $p_{\rm inj} < 1/2$. Note that at $p_{\rm inj} \gtrsim 5\%$, we are above threshold, see Fig~\ref{fig:error_location_performance}(c). 

\section{Detection-probability-based error budget}
\label{sec:defcal}

\begin{figure}[tb]
\centering
\includegraphics[width=\columnwidth]{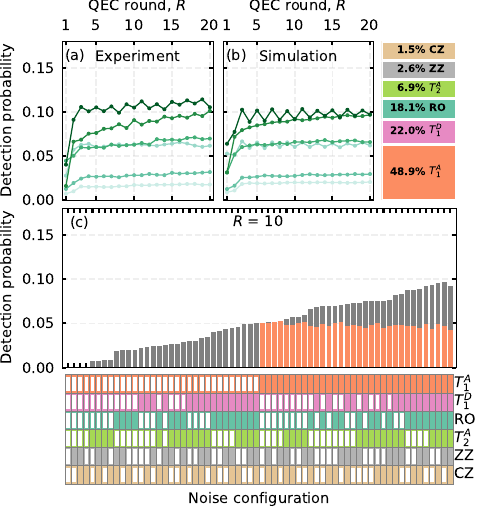}
\caption{Error budget of a single \wtwoztype stabilizer. (a) and (b) Comparison between experimental and \densitymatrix simulation results of the six characterization circuits described in Fig.~\ref{fig:DefectrateCalibration}(c). On the right side is the error budget describing a percentage contribution of each noise category to the total detection probability. (b) An ordered contribution list that shows each noise configuration at \qec round $R=10$. The largest contributing category (ancilla qubit $T_1^A$) is highlighted. Each highlighted bar represents the contribution that the category brings to the detection probability of that particular noise configuration.}
\label{fig:DefectrateErrorBudget}
\end{figure}

An error budget is useful for identifying noise contributions. In the context of \qec, the average detection probability of stabilizers is a good indication of code performance. Here we discuss a method to construct an error budget for isolated stabilizer performance by fitting a physical noise model to experimentally obtained detection probabilities. In particular, we focus on six characterization circuits described in Fig.~\ref{fig:DefectrateCalibration}(c.I-VI). By increasing the circuit complexity from (c.I) repeated ancilla measurements to (c.VI) a complete stabilizer circuit, it is possible to isolate noise sources and calculate their contribution to the total detection probability.
Note that the detectors here are no longer related to the stabilizers of a code, as there is no code. Instead, they, together with the characterization circuits are designed to be susceptible to certain specific types of errors.

The noise model used to fit the detection probabilities contains: $T_1$, $T_2$, readout assignment errors, single- and two-qubit phase errors, leakage and residual \zzc coupling. Here, the $T_1$, $T_2$ and readout assignment errors are chosen to be identical for all data qubits for simplicity. Note that the objective here is to get a qualitative sense of the noise contribution to the total detection probability. There are more direct ways of characterizing these noise parameters through standard characterization experiments. As a sanity check, we note that the parameters obtained by the fit are comparable with the experimentally obtained parameters.  
A \densitymatrix simulation is used to simulate the first 10 rounds for each of the six characterization circuits. Not all noise parameters are fitted simultaneously. First, circuits c.I, II and III are used to fit ancilla qubit $T_1$, $T_2$ and readout assignment errors. After this step, these parameters are frozen for the next step. Second, circuit c.IV is used to fit the residual \zzc phase error between data- and ancilla qubits. Third, circuits c.V and VI are used to fit data qubit $T_1$, two-qubit gate phase and leakage errors.
The root-mean square error between simulated and experimentally obtained detection probability is used as the fitness value for the given set of parameters. A COBYLA-based minimization routine is used to optimize the noise parameters with physically informed bounds.

The noise parameters obtained are grouped in categories [Fig.~\ref{fig:DefectrateErrorBudget}]. For example, the category for readout (RO) contains readout assignment- and QND infidelities, while the category for two-qubit gate (CZ) contains phase and leakage errors. The remaining noise parameters have their equally named categories.
We define a noise configuration as a subset of noise categories and simulate the complete stabilizer circuit (c.VI), obtaining an averaged detection probability.

\begin{figure}
    \centering
    \includegraphics[width=0.99\columnwidth]{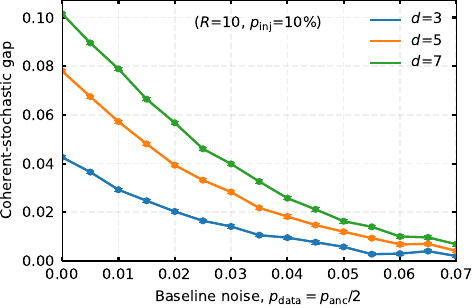}
    \caption{Free-fermion simulation of the coherent-stochastic gap [Eq.~\eqref{eq:gap}] in memory experiments with $R=10$ rounds and $p_{\rm inj} = 10\%$ injected (phenomenological) error probability as a function of the baseline noise. The baseline noise is phenomenological and specified by the data-qubit error probability $p_{\rm data}$. The ancilla-qubit error probability is chosen as $p_{\rm anc} = 2p_{\rm data}$, which is close to the ratio obtained from fitting the \disfive experimental data. The coherent-stochastic gap increases with the code distance. The memory experiment circuit is shown in Fig.~\ref{fig:fgs_simulator}(a) and run for $10^6$ samples. Note that $p_{\rm inj} = 10\%$ is above the threshold, but it leads the largest coherent-stochastic gap. }
    \label{fig:gap_vs_baseline}
\end{figure}

\begin{figure}
    \centering
    \includegraphics[width=0.99\columnwidth]{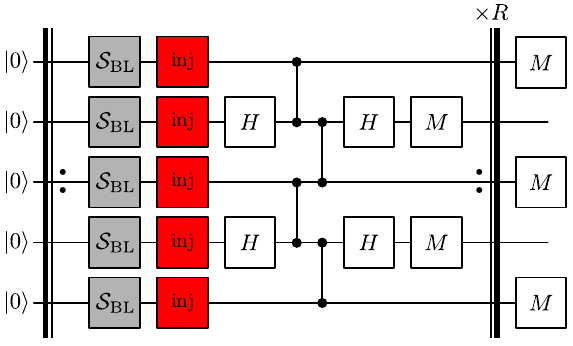}
    \caption{The noise model used in phenomenological simulations for the \disthree \repetitioncode with $R$ \qec rounds. All operations besides the explicit baseline noise channel and noise injection are assumed to be perfect. The baseline noise corresponds to the same noise channel being applied to all qubits at the beginning of each \qec round (gray boxes), with two noise strengths: one shared by all data qubits and the other by all ancilla qubits. For the model Phenomenological-$T_1$, the noise channel $\mathcal{S}_{\rm BL}$ is amplitude-damping with noise parameter $p$ as in Eq.~\eqref{eq:defp}. Twirling this noise channel gives a \bitflip noise channel with error probability $p$. 
    Errors are injected in the locations marked by red boxes and ancilla qubits are not reset at each \qec round, as in the actual experiments, see Fig.~\ref{fig:StochasticNoiseStrategy}. The model with $\mathcal{S}_{\rm BL}$ as stochastic \bitflip error can be simulated both by the \densitymatrix simulator as well as the free-fermion simulator described in Sec.~\ref{sec:sps}, hence we verify that these results are the same for \disthree.}
    \label{fig:phen-background-sim}
\end{figure}

\section{\capdensitymatrix simulations} \label{sec:dm_sim}

Here we describe the noise channels used in the circuit-level \densitymatrix simulations, as well as the choice of noise parameter values. The \densitymatrix simulations have been carried out using the \texttt{quantumsim} Python package~\cite{Quantumsim}. Table~\ref{tab:dm} contains a summary of the noise parameters.

To reduce memory consumption and speed up the simulation, low-frequency transmons that cannot leak during CZ gates have been modeled as 2-level systems (qubits), while the rest have been modeled as 3-level systems (qutrits). For the CircuitLevel-Basic simulation, all transmons have been modeled as qubits because no operation can leak the transmons outside the computational space.

\subsubsection{$R_X(\theta), R_Y(\theta)$ gates and idling}

Each single-qutrit idling is implemented by evolving the \densitymatrix state using the Lindblad Master equation with the single-qutrit Hamiltonian 
\begin{equation} \label{eq:ham_idling}
    H = \begin{pmatrix}
        0 & 0 & 0 \\
        0 & 0 & 0 \\
        0 & 0 & \alpha
    \end{pmatrix},
\end{equation}
with $\alpha$ the qubit anharmonicity, and jump operators
\begin{equation}
\begin{split}
    L_1 = \sqrt{\frac{1}{T_1}}\begin{pmatrix}
        0 & 1 & 0 \\
        0 & 0 & \sqrt{2} \\
        0 & 0 & 0
    \end{pmatrix},\, 
    L_2 = \sqrt{\frac{2}{9T_{\phi}}}\begin{pmatrix}
        1 & 0 & 0 \\
        0 & -1 & 0 \\
        0 & 0 & 0
    \end{pmatrix}, \\
    L_3 = \sqrt{\frac{8}{9T_{\phi}}}\begin{pmatrix}
        1 & 0 & 0 \\
        0 & 0 & 0 \\
        0 & 0 & -1
    \end{pmatrix},\,
    L_4 = \sqrt{\frac{2}{9T_{\phi}}}\begin{pmatrix}
        0 & 0 & 0 \\
        0 & 1 & 0 \\
        0 & 0 & -1
    \end{pmatrix},
\end{split}
\end{equation}
with $T_1$ the relaxation time, and $T_{\phi}$ the pure dephasing time~\cite{Varbanov20}, see Table~\ref{tab:dm} for our experimentally obtained device metrics.

Noisy single-qubit $R_X(\theta)$ and $R_Y(\theta)$ (shown as Hadamard in circuits) gates correspond to a sandwich of (1) idling for $\Delta t_{\rm 1Q}/2$, (2) ideal gate, and (3) idling for $\Delta t_{\rm 1Q}/2$, with $\Delta t_{\rm 1Q}$ the duration of the gate. Under this noise model and the $T_1$ and $T_2$ from Table~\ref{tab:dm}, the single-qubit gates have an infidelity of $\sim 0.1\%$, matching the experimental performance. 
Experimental characterization shows that these gates induce very little leakage, and thus we assume that they act trivially in the leaked subspace. 
The implementation of the corresponding qubit (2-level) operations can be obtained by taking the appropriate sub-matrices. 

\subsubsection{CZ gate}

The CZ gates are modeled using the \verb|quantumsim| implementation for this operation, which follows Ref.~\cite{Varbanov20, battistel2022mitigating}. This model includes leakage as an exchange between the $\ket{11}$ and $\ket{02}$ states given by
\begin{equation}
\begin{split}
    &\ket{11} \rightarrow \sqrt{1 - 4L_1} \ket{11} + e^{i\phi}\sqrt{L_1}\ket{02}, \\
    &\ket{02} \rightarrow -e^{-i\phi}\sqrt{L_1} \ket{11} + \sqrt{1 - 4L_1}\ket{02}, \\
\end{split}
\end{equation}
with $L_1$ the leakage probability ($L_1 \sim 1 - 2\%$ in our device, see Table~\ref{tab:dm}), and the leaked qutrit being the high-frequency transmon. The phase $\phi$ has been shown not to have an effect on the dynamics of leakage nor on the logical error rate~\cite{Varbanov20}, so we have set it to $\phi = \pi/2$, (the default value in \verb|quantumsim|). 
The leakage conditional phases for the $\{\ket{02}, \ket{12}\}$ subspace have been set to $\phi_{\rm stat}^{\mathcal{L}} = \pi$ and $\phi_{\rm flux}^{\mathcal{L}} = 0$, following Ref.~\cite{Varbanov20}.
We do not include leakage mobility to reduce the simulation runtime. 
If the high-frequency transmon is modeled as a qubit, the CZ operation is implemented as $\mathrm{diag}(1, \phi_{01}, \phi_{10}, \phi_{11})$, with $\phi_{ij}$ the phase accumulated by the $\ket{ij}$ state. 
Finally, the CZ gates are also sandwiched with idling to include decoherence, as done for the single-qubit gates. 

\subsubsection{\zzc coupling}

The unwanted \zzc coupling between data- and ancilla-qubits (arising from the always-on interaction mediated by the fixed-frequency coupler) is modeled as the unitary operation $U = \ket{00}\bra{00} + e^{i2\pi \zeta_{ZZ} t}\ket{11}\bra{11} + \ket{22}\bra{22}$, with $\zeta_{ZZ}$ the \zzc coupling strength [see its value in Table~\ref{tab:dm}], and with the $\ket{22}\bra{22}$ term only included when involving two qutrits. These unwanted additional unitary operations are applied continuously between pairs of coupled data- and ancilla-qubits when the two following conditions are satisfied:
\begin{itemize}
    \item the ancilla qubit is (ideally) in the XY plane (so not during measurement), and
    \item the qubit pair is not involved in a CZ gate.
\end{itemize}
The simulation simplifies the \zzc interaction with $CZ$ by only applying a single-qubit phase rotation and not considering the two-qubit phase and leakage errors introduced by a state-dependent frequency shift on one of the qubits involved in the $CZ$ operation.

\subsubsection{Measurement}

The simulated circuits only contain single-qubit $Z$-basis measurements. We model the measurement operations using the assignment matrix and the state-transfer matrix~\cite{Marques23}. The assignment matrix corresponds to $P_{\rm assign}(m_{\rm out} | \ket{s}_{\rm in})$ with $m_{out}$ the measured qutrit outcome and $\ket{s}_{\rm in}$ the input state. The state-transfer matrix corresponds to $P_{\rm QND}(\ket{s}_{\rm out} | \ket{s}_{\rm in})$ with $\ket{s}_{\rm out}$ the outgoing qutrit state. Table~\ref{tab:dm} contains the average assignment and QND fidelities in our device.
Given $P_{\rm assign}$ and $P_{\rm QND}$ of the measured qutrit, the noisy $Z$-basis measurement applied to a \densitymatrix $\rho$  is implemented as follows:
\begin{enumerate}
    \item compute the ideal probabilities of each outcome $m \in \{0, 1, 2\}$ using $P_{\rm ideal}(m) := \mathrm{Tr}(\Pi_m \rho) / \mathrm{Tr}(\rho)$, with $\Pi_m = \ket{m}\bra{m}$,
    \item sample an ideal outcome state $\ket{s}_{\rm ideal}$ from the distribution $P_{\rm ideal}$ of step 1,
    \item sample a noisy outcome $m_{\rm out}$ from the distribution $P_{\rm assign}(m_{\rm out} | \ket{s}_{\rm ideal})$ and set it as the measurement outcome,
    \item sample a noisy outgoing state $\ket{s}_{\rm out}$ from the distribution $P_{\rm QND}(\ket{s}_{\rm out} | \ket{s}_{\rm ideal})$,
    \item and project the state to $\Pi_{s} \rho \Pi_{s} / \mathrm{Tr}(\Pi_{s} \rho)$. 
\end{enumerate}

\subsubsection{Choice of noise parameters} \label{sec:choice_para}

For the most detailed \densitymatrix simulation, one could directly use the noise parameters extracted from the experimental characterization, such as a Ramsey experiment or randomized benchmarking. We have observed that these parameters are underestimating the actual noise, leading to simulated detection probabilities and a logical infidelity that are lower than the experimental ones. The reason for this discrepancy is that the characterization experiments test the performance of individual operations, without taking into account noise sources that arise when operations are executed in parallel. Evidence of this mismatch can be seen in the isolated and embedded detection probabilities from Fig.~\ref{fig:DefectratePerformance}(b). 

To match the experimental baseline noise, we have tuned the $T_1$ and $T_2$ values as follows. We have fixed the $T_1 / T_2$ ratio given by the experimental characterization and assigned the same $T_1^{\rm data}$ to all data qubits and the same $T_1^{\rm anc}$ to all ancilla qubits. Such a choice reduces the number of free parameters that need to be fitted to the experimental data without deviating a lot from the experimental values. In particular, we manually optimize $T_1^{\rm data}$ and $T_1^{\rm anc}$ to match both the experimental detection probabilities and the logical infidelity when decoding with majority voting. This same strategy is used for matching the noise parameters for the free-fermion simulation, see Sec.~\ref{sec:ff_modelling} where we motivate this choice. 
The remaining noise parameters are directly obtained from standard characterization experiments reported in Table \ref{tab:dm}.

For the CircuitLevel-Basic simulation, we cannot simply remove noise channels, as the simulated detection probabilities and logical infidelity will be lower than the experimental ones. To compensate for this underestimation of performance, we have again manually optimized the $T_1^{\rm data}$ and $T_1^{\rm anc}$ to match both the experimental detection probabilities and the majority voting performance.

\section{Scalable phenomenological free-fermion simulation}
\label{sec:sps}

This section reviews how coherent noise in an $n$-qubit \repetitioncode can be simulated efficiently classically, i.e. using at most $O(n^3)$ resources in time and space, using a mapping onto Majorana free fermion (also called fermionic linear optics or noninteracting fermions or matchgate) dynamics~\cite{Suzuki17, bravyi:FLO}. These free fermion simulation techniques have been applied to the \surfacecode in e.g.~\cite{bravyi:FLO,Marton23,pato2025}. 
For the 1D \repetitioncode, --but not for the 2D \surfacecode--, one can efficiently simulate coherent noise both on the data qubits as well as on the ancilla qubits, thus making the \repetitioncode a good testbed to study the effect of coherent errors on both data and ancilla qubits.  

In the next Sec.~\ref{sec:ff-scale}, we review how the free-fermion dynamics can be simulated in a scalable fashion. In Sec.~\ref{sec:ff_modelling}, we discuss the mapping of this free-fermion model to qubits and how one can model a \repetitioncode memory experiment being represented using free-fermion dynamics only. We then match the simulation noise parameters to the experimental ones in Sec.~\ref{sec:noreset}.

Note that we have made adaptations to the method in Ref.~\cite{Suzuki17} to capture the experimental setup. First, in the experiment, ancilla qubits are not reset at the end of each \qec round. Second, we include baseline noise from the experimental setup, but modeled in a manner that is compatible with the free fermion simulation being scalable. Third, we assume that the baseline noise is stochastic and without leakage. 

As for simulation speed and memory performance, the free-fermion simulation for $R=10$ \qec rounds for the $n=51$ \repetitioncode requires $0.392$~s (per sample) on an Apple M2 Pro processor and $\sim 16$~MB of memory. The scaling of both resources is consistent with at-most cubit scaling with the number of qubits, with a runtime of $\sim 4$~min (per sample) and memory consumption of $\sim 55$~MB for $R=10$ and $n=401$.

\begin{figure}
    \centering
    \includegraphics[width=0.99\columnwidth]{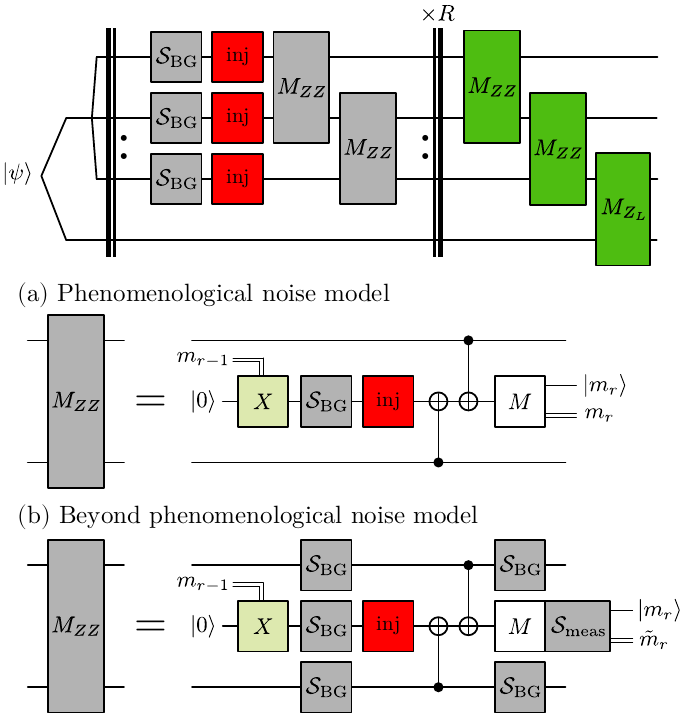}
    \caption{Memory experiment circuit and noise model simulatable by the free-fermion simulator for a \disthree \repetitioncode. There are four main differences to the circuit in Fig.~\ref{fig:phen-background-sim} to ensure that the circuit is simulatable using free-fermion dynamics. First, the ancilla qubits are not included in the circuit, but their noise and effect on the data qubits are modelled by the parity-check operators. Second, the circuit includes a noiseless extra qubit to perform the final logical measurement, $M_{Z_L}$. Third, the initial state is in a superposition, corresponding to the maximally entangled state from Eq.~\eqref{eq:ff_init_state}. Last, there are virtual conditional $X$ gates on the ancilla qubits to simulate the lack of mid-circuit resets (light green box) as described in Sec.~\ref{sec:noreset}. The noise channels $\mathcal{S}_{\rm BL}$ (gray boxes) is of the form of Eq.~\eqref{eq:general_noise_channel}. Noise channels of the form of Eq.~\eqref{eq:general_noise_channel} at the injected error locations are shown as `inj' red boxes. The noise channel $\mathcal{S}_{\rm meas}$ is described in Sec.~\ref{sec:bells_and_whistles}. }
    \label{fig:fgs_simulator}
\end{figure}

\subsection{(Non-)Unitary free fermion dynamics} \label{sec:ff-scale}

Before reformulating the qubit operations and noise channels to non-unitary free-fermionic dynamics, we first summarize the relevant properties of free fermions with regard to their simulatibility, see Ref.~\cite{Bravyi14}. The Majorana fermionic operators for $m$ fermionic modes, $\{c_i\}_{i=1}^{2m}$, satisfy $\{c_i, c_j\} = 2\delta_{ij}$, $c^{\dagger}_i = c_i$, and $c_i^2=I$. A pure (unnormalized) fermionic Gaussian state $\ket{\psi}$ is described by its covariance matrix $M \in \mathbb{R}^{2m\times 2m}$ satisfying $MM^T= I$, and its normalization $\Gamma=\bra{\psi}\psi\rangle$, with
\begin{equation}
    (M)_{ab} = \frac{i}{2\Gamma}\bra{\psi} [c_a, c_b]\ket{\psi}.
\end{equation}
The inner product between two unnormalized pure fermionic Gaussian states is given, see Fact 2 in~\cite{Bravyi14}, as 
\begin{equation}
    |\langle \phi | \psi \rangle |^2 = 2^{-m} \Gamma_{\phi}\Gamma_{\psi} \mathrm{det}(M_{\phi} + M_{\psi}). 
\end{equation}
A fermionic Gaussian operator is of the form $W= e^{\sum_{i,j} \alpha_{ij} c_i c_j}$ with complex antisymmetric matrix $\alpha_{ij}$ and it maps a pure fermionic Gaussian state to a pure fermionic Gaussian state. Note that such operator is only unitary when $\alpha=ih$ with $h$ a real anti-symmetric matrix. To characterize the effect of $W$, we can apply $W$ to half of a pure `maximally-entangled' Gaussian state~\cite{bravyi:FLO} 
\begin{equation}
    \ket{\Psi}\bra{\Psi} = \frac{1}{2^{2m}}\prod_{a=1}^{2m} \left(I + ic_a \otimes c_{2m+a} \right).
\end{equation}
The resulting pure fermionic Gaussian state $\ket{\Psi_W}=(W \otimes I) \ket{\Psi} $ has covariance matrix $M_W\in \mathbb{R}^{4m\times 4m}$ with
\begin{align}
M_W=\left(\begin{array}{cc} A & B \\ -B^T & D \end{array}\right).
\end{align}
Then, see Fact 4 in~\cite{Bravyi14}, the action of $W$ on a pure state given by $(M,\Gamma)$, gives a pure state $(M',\Gamma')$ with
\begin{align}
    M' &= A-B(M-D)^{-1} M B^T, \\
    \Gamma' &= \Gamma_W \Gamma \sqrt{\mathrm{det}(M-D)} \label{eq:traceff},
\end{align}
with $\Gamma_W=\bra{\Psi_W}\Psi_W\rangle$, using the matrices $A,B,D$.
Hence, any dynamics in which (1)~one starts with a pure Gaussian fermionic state, (2)~applies a TCP map of which the Kraus operators are Gaussian operators (which includes Gaussian unitary dynamics), and (3)~possibly finally measures elements of the final covariance matrix, can be classically simulated by handling the efficiently-computable updates to the $2m \times 2m$ covariance matrix. The computation of the trace in Eq.~\eqref{eq:traceff} is used to evaluate the probability for distinct outcomes for the TCP map, so that one can appropriately sample from the distribution and continue with one of the pure Gaussian trajectories.

\subsection{Mapping and modeling} \label{sec:ff_modelling}

The mapping of an $n$-qubit \bitflip \repetitioncode memory experiment to non-unitary free-fermion dynamics uses the Jordan-Wigner transformation to map qubits to fermions. For technical reasons, see below, the total number of qubits will be $m=n+1$. We identify $c_{2i-1} = (\prod_{j=1}^{i-1}X_j) Z_i$ and $c_{2i} =(\prod_{j=1}^{i-1}X_j) Y_i$. This choice leads to bit flips and \repetitioncode parity checks being quadratic operators in the $\{c_i\}$:
\begin{align} \label{eq:JW}
    X_i = -ic_{2i}c_{2i-1}, \; Z_i Z_{i+1} = -i c_{2i+1} c_{2i}.
\end{align}
Hence, a coherent rotation $R_X(\theta)$ corresponds to a Gaussian unitary transformation. In fact, a more general \bitflip noise channel $\mathcal{E}_{\rm mix}(\rho)$ corresponding to a mixture of fully-coherent and stochastic $X$-type noise is also a Gaussian operator:
\begin{equation} \label{eq:general_noise_channel}
    \mathcal{E}_{\rm mix}(\rho) = c\mathcal{E}_{\rm coh}(\rho) + (1-c)\mathcal{E}_{\rm stoch}(\rho),
\end{equation}
with the coherent rotation $\mathcal{E}_{\rm coh}$ and \bitflip channel $\mathcal{E}_{\rm stoch}$ defined in Eq.~\eqref{eq:noise_channels}. Here $0 \leq c \leq 1$ is the noise coherence, with $c=0$ stochastic and $c=1$ fully-coherent noise, and the Kraus operators for this two-outcome $\pm$ channel are
\begin{equation}   
A_{\rm mix}^{+} = \sqrt{\frac{1 + c}{2}}R_X(\theta) \;\;\text{and}\;\; A_{\rm mix}^{-} = \sqrt{\frac{1 - c}{2}}R_X(-\theta). 
\label{eq:kraus}\end{equation}

The Kraus operators of a perfect \wtwoztype parity measurement, i.e. $K_{s=0,1}=\frac{1}{2}(I+(-1)^{s} ZZ)$ correspond to a (non)-unitary Gaussian operator. Hence, in modeling the memory experiment dynamics, the ancilla qubits are not explicitly included in the fermion representation: this keeps the simulation free-fermionic. It can also be noted that amplitude damping noise, which is experimentally relevant, does not have Gaussian Kraus operators as its Kraus operators involve all qubit Paulis, $X$, $Y$ and $Z$.

The memory experiment of an $n$-qubit \bitflip \repetitioncode under phenomenological noise involves (1) initializing the data qubits in $\ket{b_1, ..., b_n}$ with $b_i \in \{0,1\}$; (2) measuring the stabilizers of the code, corresponding to $\{Z_i Z_{i+1}\}_{i=1}^{n-1}$; (3) measuring the logical qubit in the $Z$-basis, with $Z_L = Z_n$ (say), and (4) including the noise in these processes. In Fig.~\ref{fig:phen-background-sim}(a) we show a simple phenomenological model of baseline noise plus injected noise at the start of each \qec round (the two channels can be lumped together in the simulation). For this model we assume that the ancilla qubits are reset between rounds and that there is no baseline noise. 

To model injected noise with $\mathcal{E}_{\rm mix}$ in Eqs.~\eqref{eq:general_noise_channel} and \eqref{eq:kraus} on the ancilla qubits, we can modify the CP maps corresponding to syndrome outcome $s=0,1$, i.e. one has for the measurement on data qubits $i$ and $i+1$:
\begin{align} \label{eq:Kraus_parity}
    \mathcal{S}_{s}(\rho)&=K_{s,+}\rho 
    K_{s,+}^{\dagger}+K_{s,-}\rho K_{s,-}^{\dagger} \notag \\
        K_{s, \pm} &= \frac{1}{2} \sqrt{\frac{1\pm c}{2}} \left( 1 + (-1)^s e^{\pm i \theta} Z_i Z_{i+1} \right),
\end{align}
where $K_{s,\pm}\equiv K_{s,\pm}(\theta,c)$ is a Gaussian operator as $Z_iZ_{i+1}$ is quadratic in Majorana operators under the Jordan-Wigner transformation. Note that to include a stochastic \bitflip noise baseline on the ancilla qubits, one can simply associate a flipped $s\rightarrow \overline{s}$ outcome with the map $\mathcal{S}_s$ with probability equal to the \bitflip probability. One includes baseline plus injected noise on the data qubits as in Fig.~\ref{fig:phen-background-sim}(a) by applying these unitary Gaussian operators prior to each \qec measurement round.

For the scalable simulation we also need the input state, i.e. $\ket{0_L}=\ket{0}^{\otimes n}$ or $\ket{1_L}=\ket{1}^{\otimes n}$ or another $n$-qubit bitstring, to be Gaussian. In addition, the logical measurement projectors, i.e. $(1 \pm Z_n)/2$, need to be Gaussian projectors. This issue is solved in Ref.~\cite{Suzuki17} by adding a reference qubit and initializing the $m=n+1$ qubit system in 
\begin{equation} \label{eq:ff_init_state}
    \ket{\psi}=\frac{\ket{0_L, 0} + \ket{1_L, 1}}{\sqrt{2}},
\end{equation}
which is a fermionic Gaussian state. It is a unique eigenstate of commuting Gaussian projectors onto $Z_iZ_{i+1}=+1$ and $X_1X_2\ldots X_{n+1}=+1$ where the all $X$-string is equivalent to the quadratic operator $\propto i c_1 c_{2n+2}$ via Eq.~\eqref{eq:JW}.
As for the final measurement, note that the non-unitary free-fermion simulation does not include individual single-qubit measurement in the $Z$-basis as depicted in Fig.~\ref{fig:phen-background-sim} (or as they are happening in the experiment). However, one only uses these measurements to construct a set of final \zzc syndromes and the outcome of the logical measurement.
Hence, one can model this as measuring the \zzc checks without noise which is a Gaussian operator (perfect $M_{ZZ}$ in the green boxes in Fig.~\ref{fig:fgs_simulator}), while additional measurement noise could be added by flipping the \zzc outcomes.
The logical information is extracted by first applying a noise-free projective measurement with Gaussian Kraus operators at the end of the simulation (perfect bare $M_{Z_L}$ in the green box in  Fig.~\ref{fig:fgs_simulator}), i.e. we apply the map with Kraus operators
\begin{equation}
    K_{Z_L=l} = \frac{1 + (-1)^l Z_n Z_{n+1}}{2},
\end{equation}
with $l\in\{0, 1\}$, where qubit $n+1$ is the reference qubit. Since the reference qubit is noise-free, the outcome $l$ of this measurement will be 0 if the noise on the data qubits has not flipped $Z_L = Z_n$ and $1$ otherwise. To include decoding in the scalable free fermion simulation, the syndrome measurement outcomes ${\bf s}$ of all \qec rounds, ---including that of the final noise-free measurement---, are passed to a decoder which decides on a correction bit $b=0,1$. The final outcome of a run of the simulation is $b\oplus l$. The run is labeled as failure when $b \oplus l\neq 0$ and the logical error probability $P_L$ is estimated as the total number of failed runs versus the total number of runs. Note that unlike in the actual experiment or other simulations, we estimate the logical failure of both $\ket{0_L}$ and $\ket{1}_L$ simultaneously, hence corresponding to the average logical failure probability for $\ket{0}_L$ and $\ket{1}_L$. The decoder that we use is described in Sec.~\ref{sec:dec} and it does not use any information about injected or baseline noise.

\subsubsection{No reset modification} \label{sec:noreset}

We have adapted the method from Ref.~\cite{Suzuki17} described in the previous section to not include ancilla-qubit resets during the \qec rounds. First,consider the outcome of the \zzc checks in Eq.~\eqref{eq:Kraus_parity} to be $m$ for measurement instead of $s$. Then, we insert in the model for the next \qec round an $X$ gate depending on $m=0$ or $m=1$ after the ancilla qubit is initialized in $\ket{0}$, see Fig.~\ref{fig:fgs_simulator}(a) for round r. As this conditional Pauli $X$ commutes with the baseline and injected noise, it can be propagated forward and simply flips the outcomes $m \rightarrow \overline{m}$ of its Kraus operators (which is tracked in the simulation). As is well known, in the presence of no reset, the syndrome, ---the eigenvalues of \zzc checks---, are then constructed as $m_i \oplus m_{i+1}$ and passed onto the decoder.

\subsubsection{Matching the experimental performance} \label{sec:ff_matching}

We have used the phenomenological noise model depicted in Fig.~\ref{fig:fgs_simulator}(a) for all simulations trying to capture the performance of the experiment.

In principle, a more detailed simulation option, see also Sec.~\ref{sec:bells_and_whistles}, would be to pick parameters to match the performance of a single parity-check measurement, which is something we tried initially. However, for our device, isolated parity tomographic data ---from which noise information could be potentially obtained--- does not match the full experimental performance of this device very well due to \zzc cross-talk and the need for flux pulses to park qubits during CZ gates, see Sec.~\ref{sec:charbackground}. Simulation using this isolated parity tomographic data predicts lower logical error and detection probabilities and thus we have not included it in our work. 

\begin{figure}[tb]
    \centering
    \includegraphics[width=0.99\columnwidth]{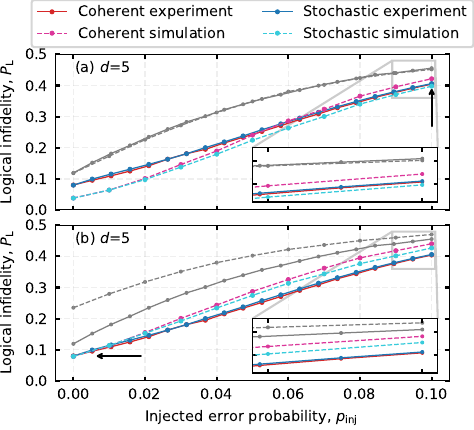}
    \caption{Logical infidelity for the experimental and (free-fermion) simulated \disfive \repetitioncode under injected coherent and stochastic noise. The simulation configuration is the same as in Fig.~\ref{fig:ErrorInjectionResults} for panel (a), but with different $p_{\rm data}$ and $p_{\rm anc}$ for panel (b). The two noise parameters have been matched to the experimental detection probabilities and the logical performance when decoded with majority voting for panel (a), resulting in the simulation matching the logical performance at $p=10\%$ (see arrow indication), and with MWPM for panel (b), resulting in the simulation matching the logical performance at $p=0\%$ (see arrow indication). Even though the baseline performance is different, in both cases the simulation predicts a coherent-stochastic gap. The simulated number of samples per point is $10^5$. 
    }
    \label{fig:ff_matching}
\end{figure}

For our phenomenological modeling, we assume that all data qubits undergo stochastic \bitflip noise at the beginning of each \qec round with physical error probability $p_{\rm data}$. For the ancilla qubits, we take the same stochastic \bitflip noise channel but with error probability $p_{\rm anc}$. Having only two free noise parameters makes the process of matching to the experimental performance fairly simple. We manually optimize the two parameters to match both the total average experimental detection probabilities and the logical performance when decoding with majority voting (which does not use parity check data obtained in the rounds), where we allow to pick different $p_{\rm data}$ and $p_{\rm anc}$ for \disthree versus \disfive. The reason for matching both detection probabilities and logical performance is to ensure that the optimization problem is not under-constrained: if we only matched the detection probabilities, there are different combinations of $(p_{\rm data},\,p_{\rm anc})$ that give the same detection probabilities but that have different logical performance. The optimized values for $p_{\rm data}$ and $p_{\rm anc}$ are 2.9\% and 3.5\% respectively for \disthree, and 3.3\% and 7.3\% for \disfive. The \disfive code has higher baseline noise due to the longer duration of the \qec round, among others.

As it turns out, fitting the detection probabilities and majority voting leads to a good match at a high injected error probability, but to a bad one at small injected error probabilities [see Fig.~\ref{fig:ErrorInjectionResults}]. In particular, the simulation predicts better baseline performance than what is achieved experimentally. This could potentially lead to a bias in simulation as it underestimates the baseline noise, possibly leading to a coherent-stochastic gap. To test this bias, we have also manually optimized $p_{\rm data}$ and $p_{\rm anc}$ to match both the total average experimental detection probabilities and the logical performance when decoding with MWPM. The new parameters match the baseline performance and still show a coherent-stochastic gap [Fig.~\ref{fig:ff_matching}]. 

\begin{figure}[tb]
    \centering
    \includegraphics[width=0.99\columnwidth]{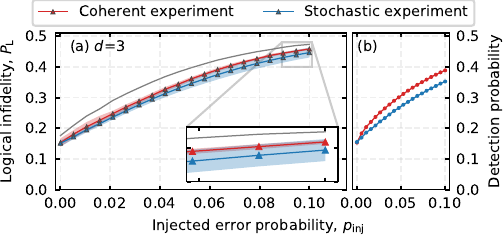}
    \caption{Additional dataset executing the same experiment as Fig.~\ref{fig:ErrorInjectionResults}, but under different calibration conditions.
    Shows a difference in detection probability under coherent and stochastic noise injection.
    However, the uncertainty in the estimation of the stochastically injected logical infidelity is larger than the expected coherent-stochastic gap. Uses bootstrapping method with 10 sample groups to estimate uncertainty.
    }
    \label{fig:ExtraResults}
\end{figure}

\subsection{Beyond phenomenological noise modeling} \label{sec:bells_and_whistles}

In this section, we briefly discuss a more general scalable free-fermion model that has many noise parameters, but we have not used it in simulations so far due to the large number of parameters to match with experimental results. It may be possible to use machine learning techniques, i.e. reinforcement learning to find the parameters which give the best match with experimental data. 

In addition, one could replace some non-free fermion operations by fermionic magic resources and use non-scalable fermionic simulation techniques~\cite{hebenstreit,Dias_2024,Reardon_Smith_2024}. It is an open question whether these could be sufficiently fast and/or outperform Clifford or beyond-Clifford circuits in accurately capturing experimental performance. 

The more general free-fermion noise model is depicted in Fig.~\ref{fig:fgs_simulator}(b). First of all, all noisy baseline channels $\mathcal{S}_{\rm BL}$ which are all of the form in Eq.~\eqref{eq:general_noise_channel}, including the incoming noise on the data qubits in Fig.~\ref{fig:fgs_simulator}, can be qubit-specific, round-specific, and circuit location-specific. Note that two subsequent applications of $\mathcal{E}_{\rm mix}$ in Eq.~\eqref{eq:general_noise_channel}, with possibly different parameters, can always be lumped together since
\begin{align}
    \mathcal{E}_{\rm mix}(c_1,\theta_1) \circ \mathcal{E}_{\rm mix}(c_2,\theta_2)= \mathcal{E}_{\rm mix}(c_3,\theta_3).
    \label{eq:lump}
\end{align}
This holds as $\mathcal{E}_{\rm mix}(c, \theta)$ only affects off-diagonal elements in the $X$-basis, i.e. $\mathcal{E}_{\rm mix}(c, \theta)(\ket{+}\bra{-})=\lambda(c,\theta)(\ket{+}\bra{-})$ with $\lambda(c,\theta)$ obtained as a convex combination (depending on $c$) of $e^{i\theta}$ and $e^{-i\theta}$ which can represent any point in the unit disk.

For every $M_{ZZ}$ measurement, there can be incoming baseline noise and outgoing baseline noise (sometimes they can be lumped together as in Eq.~\eqref{eq:lump}). The channel $\mathcal{S}_{\rm meas}$ in Fig.~\ref{fig:fgs_simulator}(b) is an additional classical (possibly asymmetric) \bitflip channel which makes the outgoing ancilla qubit state $\ket{m}$ differ from the read-out bit $\tilde{m}$. Although $\mathcal{S}_{\rm meas}$ does not allow to represent an arbitrary noisy measurement, it allows including in simulation most of the noise sources present in the experiments, e.g., qubit relaxation and/or excitation during measurement. 

\end{document}

%% file: Preamble/packages.tex
\usepackage[makeroom]{cancel}
\usepackage{dcolumn}
\usepackage{makecell}
\usepackage[utf8]{inputenc}
\usepackage[T1]{fontenc}
\usepackage{graphicx}
\usepackage{multirow}
\usepackage{amsmath}

\usepackage{soul}
\usepackage{verbatim}
\usepackage{bm}
\usepackage{outlines}
\usepackage{siunitx}
\usepackage[space]{grffile}
\usepackage[Symbolsmallscale]{upgreek}
\usepackage[dvipsnames]{xcolor}
\usepackage{xr}
\usepackage{color}
\usepackage{tikz}
\usepackage{mathtools}
\usetikzlibrary{quantikz2}
\usepackage{xspace}

\newcommand\myshade{85}

\colorlet{mylinkcolor}{violet}
\colorlet{mycitecolor}{YellowOrange}
\colorlet{myurlcolor}{Aquamarine}
\usepackage{hyperref}
\hypersetup{
  linkcolor  = mylinkcolor!\myshade!black,
  citecolor  = mycitecolor!\myshade!black,
  urlcolor   = myurlcolor!\myshade!black,
  colorlinks = true,
}
\usepackage[capitalize]{cleveref}
\usetikzlibrary{calc, positioning} 

\usepackage{amssymb,placeins}

%% file: Bibliography/bibliography_main.tex
%